\documentclass[twocolumn]{aa} 

\usepackage[varg]{txfonts}   
\usepackage{graphicx}        
\usepackage{amsmath}         
\usepackage{natbib}          
\usepackage{hyperref} 
\usepackage{tabularx}
\usepackage{caption}
\usepackage{subcaption}
\usepackage{rotating}
\usepackage{float}
\usepackage{placeins}
\hypersetup{
    colorlinks=true,
    linkcolor=blue,
    citecolor=blue,
    urlcolor=blue,
}

\begin{document}

\title{Asteroseismic properties of young pulsators in the Perseus Molecular Cloud}
\subtitle{Discovery of period-spacing patterns in HD~281229}
\author{
    Saakshi Wadhwa\inst{1,2},
    Dario J. Fritzewski\inst{1}, Gang Li\inst{2}, and Simon J. Murphy\inst{2}
}
\authorrunning{Wadhwa et al.}

\institute{
    Institute of Astronomy, KU Leuven, Celestijnenlaan 200D, 3001 Leuven, Belgium \and 
    Centre for Astrophysics, University of Southern Queensland, Toowoomba, QLD 4350, Australia \\
    \email{saakshi.wadhwa@unisq.edu.au, dario.fritzewski@kuleuven.be}
}


\abstract
{Very young pulsating stars, including pre-main sequence pulsators, provide a rare opportunity to probe stellar interiors and investigate rotation during the earliest phases of stellar evolution. However, they remain relatively unexplored, particularly in the intermediate-mass regime.}
{We aim to identify and characterise young pulsating stars in the Perseus Molecular Cloud using photometric data from the Transiting Exoplanet Survey Satellite (TESS).}
{We compiled \emph{Gaia}-based membership catalogues for the Perseus Molecular Cloud, applied selection criteria to isolate young intermediate-mass stars, and extracted light curves from TESS full-frame images. Variability classification was performed using Lomb--Scargle periodograms and iterative pre-whitening. For confirmed pressure-mode pulsators, we estimated the frequency of maximum pulsation power, $\nu_{\mathrm{max}}$. For gravity-mode candidates, we searched for period-spacing patterns and fit them using the traditional approximation of rotation.}
{From the 148 membership-selected targets, we identified eight pressure-mode pulsators and two objects showing possible gravity-mode pulsations. We also analysed the previously known $\delta$~Sct star V705~Per, bringing the total sample to 149 targets. One of the gravity-mode pulsators, HD~281229, is the youngest star known to date with a detected and modelled period-spacing pattern. By simultaneously fitting the dipole and quadrupole period-spacing patterns, we estimate an asymptotic period spacing of \(\Pi_0=4776\pm400\,\mathrm{s}\) and a near-core rotation frequency of \(f_{\mathrm{rot}}=1.317\pm0.05\,\mathrm{d}^{-1}\).}
{These results highlight the potential of asteroseismology to probe the internal structure and rotation of very young stars. The detected period-spacing pattern in HD~281229 shows that gravity-mode asteroseismology can provide meaningful constraints on the properties of stars at early evolutionary stages.}


\maketitle
\section{Introduction}\label{1}
Very young stars near the zero-age main sequence (ZAMS), both still pre-main-sequence (PMS) or already on the main sequence, provide an important opportunity to study stellar structure at early evolutionary stages. This phase sets the conditions for subsequent stellar evolution. However, many aspects of this early evolution of intermediate-mass stars ($1.3 < M/M_\sun<8$) remain poorly understood, and uncertainties in the internal structure of young stars can therefore propagate into models of later evolutionary stages \citep{Steindl22}. Developing a clearer picture of this phase is thus essential for improving our understanding of stellar evolution as a whole.

In the classical picture, PMS stars contract with approximately constant mass along the Hayashi and Henyey tracks in the Hertzsprung--Russell diagram \citep{Hayashi,henyey}. However, this framework assumes a
fully assembled star and neglects the role of ongoing mass accretion during the formation. \citet{Steindl22} instead modelled this evolution from a protostellar seed that grows through disc-mediated accretion. The resulting evolutionary paths depend on the accretion history and differ substantially from the classical Hayashi and Henyey framework, leading to differences in the internal structure of stars at young ages. Asteroseismology provides a direct means of probing these internal properties. Intermediate-mass stars are particularly valuable in this context because they can exhibit pulsational variability during the PMS phase, allowing their interior structure to be studied through asteroseismology \citep{Zwintz2022}.

The first discoveries of pulsating PMS stars date back to the early 1970s, when \citet{Breger1972} identified $\delta$~Scuti-type variability in two young stars, V~588~Mon and V~589~Mon, located in the $\sim$3 Myr old open cluster NGC~2264. However, it took another two decades before follow-up detections emerged by \citet{Kurtz1995} of a $\delta$~Sct pulsation in the Herbig Ae star, HR~5999, which sparked renewed interest in PMS asteroseismology. Subsequent theoretical calculations of radial and non-radial pulsation modes provided the framework needed to interpret the observed frequency spectra and relate them to the internal structure and evolutionary state of PMS stars \citep{ruoppo,murphy21,Zwintz2022}.

Asteroseismology uses stellar pulsations to infer internal stellar structure and dynamics. In stars with convective cores and radiative envelopes, two types of oscillation modes are commonly observed, distinguished by their dominant restoring force. Pressure modes (p modes), where pressure acts as the restoring force, primarily probe the outer stellar layers and are generally observed at frequencies above approximately 5\,d$^{-1}$. Gravity modes (g modes), restored by buoyancy, are sensitive to stellar structure in the deep radiative interior and are typically detected at lower frequencies of about 0.3--5\,d$^{-1}$ \citep{Aerts2010}.

Among intermediate-mass stars, $\delta$~Scuti ($\delta$~Sct) stars exhibit low-order p-mode oscillations driven by the $\kappa$ mechanism operating in the He\,II ionisation zone. Gravity-mode pulsators span a broader range of stellar parameters. Cooler A- and F-type g-mode pulsators have traditionally been classified as $\gamma$~Doradus ($\gamma$~Dor) stars, whose high-order non-radial g modes are driven by convective flux blocking at the base of the envelope convection zone \citep{dupret,Aerts2010}. At higher temperatures, B-type g-mode pulsators are classically identified as slowly pulsating B (SPB) stars, in which high-order g modes are excited by the $\kappa$ mechanism associated with the iron-group opacity bump \citep{Aerts2010}. However, recent \emph{Gaia} and TESS studies have shown that g-mode pulsators are not confined to these classical instability regions, but occur across a much broader part of the Hertzsprung--Russell diagram, including stars between and beyond the traditional $\gamma$~Dor and SPB domains \citep{gaiacollab,Hey2024}. Some intermediate-mass stars exhibit both p- and g-mode oscillations and are therefore classified as hybrid pulsators \citep{grigahcene2010,xiong2016,kliapets2025}.

In p-mode pulsators, the two seismic parameters large frequency separation, $\Delta\nu$ \citep{bedding20} and the frequency of maximum pulsation power, $\nu_{\mathrm{max}}$ \citep{Murphy2024} are of particular interest. These quantities have been studied in main-sequence stars, but their behaviour in PMS stars remains poorly constrained. Based on stellar evolution models, \cite{murphy23} and \cite{Gautam2026} explored the relationship between $\Delta\nu$ and other asteroseismic indices with mass, age, metallicity and rotation for both MS and PMS stars.

In contrast, g-mode pulsators such as $\gamma$ Dor stars exhibit high-radial order, non-radial gravity modes that can form quasi-regular period-spacing patterns in the asymptotic regime \citep{shibahashi1979,tassoul1980}. In the simplest limit, modes of the same angular degree and consecutive radial order are approximately equally spaced in period, while rotation and chemical gradients introduce systematic deviations from this behaviour \citep{bouabid2013,ouazzani2017}. Thus, when detected, these period-spacing patterns provide direct constraints on the near-core rotation rate, $f_{\mathrm{rot}}$, and on the asymptotic period spacing, $\Pi_0$, which is sensitive to the stratification in the deep stellar interior \citep{miglio2008}.

Stellar rotation is a key ingredient of (early) stellar evolution, because it affects the internal structure, mixing, and angular momentum transport of stars \citep{aerts24}. Rotation measurements in young populations provide a diagnostic of how these processes operate during the earliest stages of stellar evolution. Photometric time-series observations can be used to measure the surface rotation frequency from periodic brightness variations caused by rotational modulation, for example from starspots or other surface inhomogeneities \citep[e.g.][]{Fritzewski_2016,rebull2020}, while asteroseismology can also probe the near-core rotation rate through period-spacing patterns for g modes. Fitting of p mode frequencies and comparison with spectroscopic $v\sin i$ can help to constrain the inclinations (Gautam in prep.). Together, these diagnostics make it possible to characterise rotation at different depths in very young stars and to place the measured rotation properties in the broader context of early angular momentum evolution.

Since the early discoveries, space-based photometric missions have significantly expanded the sample of known PMS pulsators. Observations from the Microvariability and Oscillations of STars (MOST) \citep{most}, Convection, Rotation and planetary Transits (CoRoT) \citep{corot}, \textit{Kepler}/K2 \citep{kepler} missions, and more recently Transiting Exoplanet Survey Satellite (TESS) \citep{2014ricker} have revealed numerous young pulsating stars in clusters and star-forming regions \citep{Zwintz2009, Zwintz2013, Mellon,murphy21, Steindl2021a}. These observations have uncovered not only p-mode pulsators but also g-mode and hybrid pulsators. Additionally, \citet{Steindl2021a} detected tidally perturbed PMS stars and \citet{mullner} detected candidates for solar-like oscillations.

Despite these advances, the number of confirmed PMS pulsators remains limited. As compiled by \citet{Zwintz2022}, more than 100 pressure mode PMS pulsators have been identified, but fewer than a dozen gravity mode pulsators and only four hybrid pulsators are currently known. Crucially, only one PMS g-mode pulsator with a period-spacing pattern is reported in the literature \citep{Ketzer2020}. This highlights the need for further systematic searches and detailed follow-up studies in young clusters and star-forming regions.

The Perseus Molecular Cloud (PMC) is one of the nearest active star-forming regions, located at a distance of approximately 300~pc \citep{ortizleon}. It hosts the well-known embedded clusters IC~348 and NGC~1333, which trace different evolutionary stages, with NGC~1333 representing one of the youngest populations in the region and IC~348 being slightly more evolved \citep{Luhman_2016,Pavlidou2021}. In addition to these clusters, \emph{Gaia}-based studies have revealed a broader population of young stellar groups distributed across the cloud, including Autochthe, Alcaeus, Heleus, Electryon, Mestor, and Gorgophone \citep{Pavlidou2021,Olivares2023}. Recent age estimates for the main Perseus groups suggest an age sequence across the region, with NGC~1333 and Autochthe at about 1~Myr, IC~348 at about 2~Myr, Heleus at 2--3~Myr, Mestor at 3--4~Myr, Electryon at 4--5~Myr, and Alcaeus at 5--10~Myr \citep{Pavlidou2021,Olivares2023,pavlidou26}. This diversity in age makes the PMC a particularly valuable environment for probing stellar variability across different stages of early evolution.

In recent years, asteroseismology of stellar clusters has become an important approach for investigating stellar interiors and evolutionary processes \citep[e.g.][]{bedding23,palak,fritzewski24,li2025,Fritzewski_2026,wang2026}. Extending such analyses to very young star-forming regions is particularly valuable, as it enables the study of pulsations during the earliest phases of stellar evolution.

The aim of this study is to identify and characterise young pulsators including PMS stars in the PMC using photometric observations from the TESS mission. By applying frequency analysis and asteroseismic techniques to both p-mode and g-mode candidates, we aim to extract global pulsational parameters, identify potential period-spacing patterns, and investigate the internal properties of young stars. The structure of this paper is as follows. In Sect.~\ref{2}, we describe the target selection, membership compilation, and TESS observations used in this work. Section~\ref{3} outlines the light-curve extraction, cleaning, variability classification, and pre-whitening procedure. In Sect.~\ref{4}, we present the asteroseismic analysis of the identified p-mode and g-mode pulsators. Finally, Sect.~\ref{5} summarises the main results and discusses their implications for asteroseismology near the ZAMS.

\section{Data and Sample Selection}\label{2}

\subsection{Membership}
We compiled our initial target list from the two \emph{Gaia}-based membership studies of the PMC by \citet{Pavlidou2021} and \citet{Olivares2023}, both of which used clustering in astrometric space to identify young stellar groups within the PMC.  Together, \citet{Pavlidou2021} and \citet{Olivares2023} identified 1677 members, from which we selected a subsample for this study.

\subsection{Sample Selection Criteria}
We cross-matched the full membership list with \emph{Gaia} DR3 data to obtain stellar parameters and applied three selection criteria to isolate young intermediate-mass stars. Firstly, stars fainter than $G=16$\,mag were excluded to ensure sufficient photometric signal-to-noise in the TESS full-frame images (FFIs). Secondly, for stars with available \emph{Gaia} GSP-Phot effective temperatures, we required $T_{\mathrm{eff}}>6500$\,K, corresponding approximately to spectral types earlier than mid-F\footnote{\label{mamajekfn}\url{https://www.pas.rochester.edu/~emamajek/EEM_dwarf_UBVIJHK_colors_Teff.txt}} \citep{mamajek}. GSP-Phot temperatures were available for 56\% of the membership list. Stars without these estimates were retained at this stage and assessed using the colour criterion below. Thirdly, we applied a colour cut of $(G_{\rm BP}-G_{\rm RP})<2.587$\,mag to all remaining stars. This limit corresponds to the intrinsic \emph{Gaia} colour of an F5-type star, $(G_{\rm BP}-G_{\rm RP})_0=0.587$\,mag\footnotemark[\getrefnumber{mamajekfn}], combined with an average colour excess of $E(G_{\rm BP}-G_{\rm RP})\sim2.0$\,mag. Because the PMC exhibits differential reddening, this value was adopted as an empirical average based on the \emph{Gaia} colour excesses of the sample.

\begin{figure}
    \centering
    \includegraphics[width=1\linewidth]{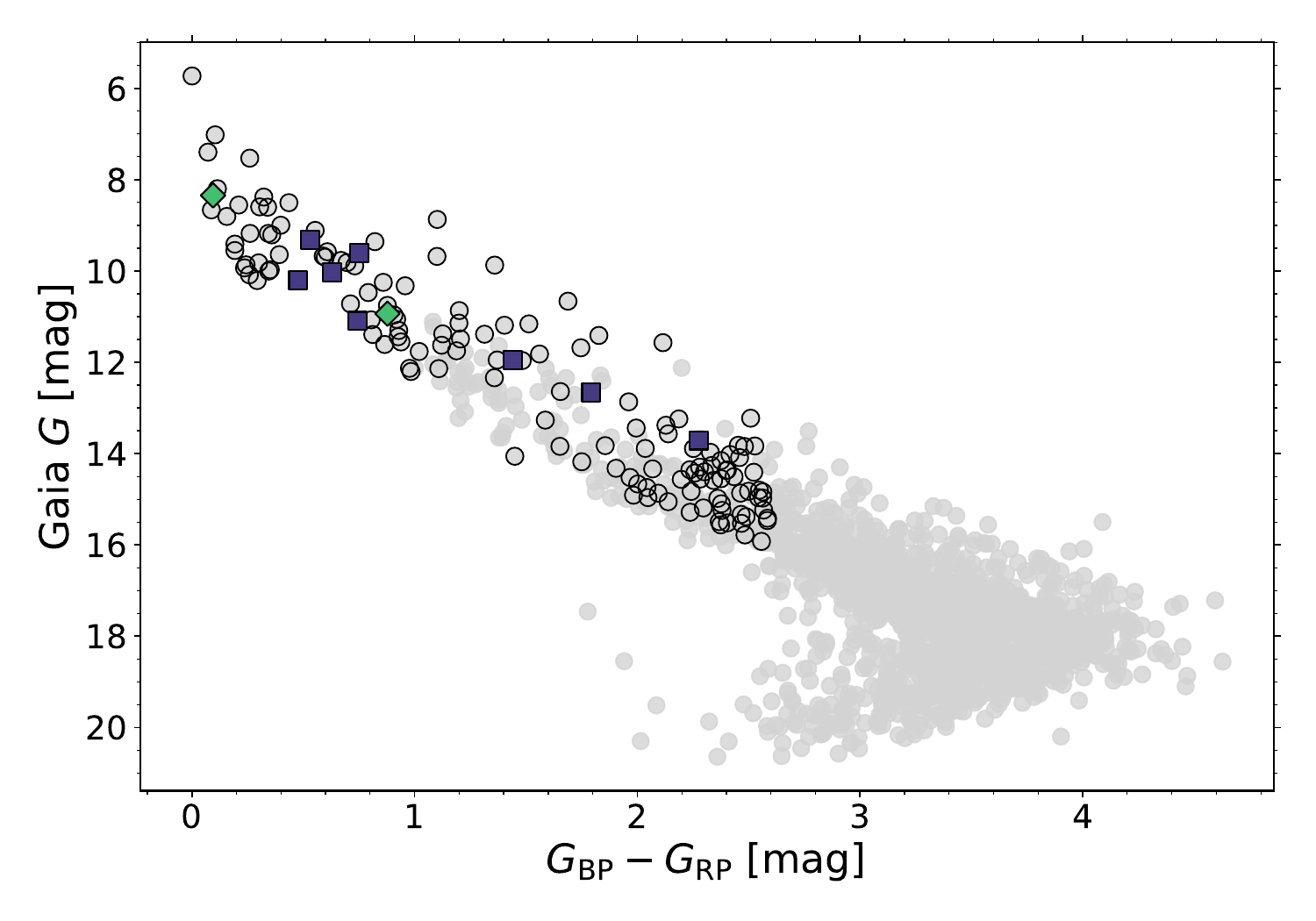}
    \caption{Colour–magnitude diagram of the Perseus Molecular Cloud. Light grey points represent the full member catalogue based on \citet{Pavlidou2021} and \citet{Olivares2023}, while black circles mark our targets. The diamonds and squares indicate the g-mode and p-mode pulsators identified in this study, respectively.}
    \label{fig:cmd2}
\end{figure}

This filtering yielded 148 targets. Additionally, we included V705~Per, a previously identified $\delta$~Sct star in the region \citep{ripepi} that was not part of the membership-based target list (see Sec. \ref{411}). These stars are spatially distributed across all subgroups and fall predominantly along the hot, blue end of the CMD, consistent with intermediate-mass candidates as shown in Fig.~\ref{fig:cmd2}.

\subsection{TESS observations and sector coverage}
The PMC was observed mainly during Year 4 of the TESS mission, when the satellite conducted a dedicated survey of the ecliptic plane. FFI data from sectors 41–44 were used as the primary dataset; these sectors provide a 10-minute cadence, ideal for capturing variability in intermediate-mass pulsators. Where available, additional observations from TESS sectors 58, 70, 71, and 72 were included for each target. Sector 18, observed at 30-minute cadence, was excluded due to insufficient time resolution for detecting high-frequency pulsations typical of $\delta$~Sct stars.

\section{Photometry and Variability Classification}
\label{3}

\subsection{Light curve extraction and cleaning}
We extracted light curves for the selected sample of $\sim$150 young stars from TESS FFIs using \texttt{tessutils} \citep{garcia}, a custom package built on top of \texttt{lightkurve} \citep{Lightkurve2018} and optimised for asteroseismology. The pipeline automates several key steps in time-series photometry from TESS FFIs, including the download of FFI cut-outs through TESScut \citep{tesscut}, the construction of optimal photometric apertures, and correction for background and contamination. In addition to raw and detrended light curves, \texttt{tessutils} generates stitched multi-sector time series. This approach proved particularly effective in crowded fields and for stars observed in multiple sectors \citep{li24,fritzewski24,Fritzewski_2026}.

We supplemented this fully automated workflow with selected manual intervention to optimise the final light curves. The sector-wise light curves were visually inspected for instrumental artifacts, flux discontinuities, and episodes of increased background noise. A small number of affected intervals were manually masked (see Table~\ref{masked_intervals}), after which the remaining long-term trends and sector offsets were corrected by fitting a third-degree polynomial to each light curve. The detrended flux was then normalised and stitched across sectors to produce a continuous time series.

While \texttt{tessutils} successfully extracted light curves for the majority of the sample, a subset of targets required additional steps. In 55 cases, automated aperture placement failed due to strong contamination from nearby stars, non-isolated targets, or instances where the star lay outside the initial detection mask. For 30 of these stars, modifying pipeline settings, such as increasing the mask size or adjusting aperture thresholds, allowed successful extraction. For the remaining 25 stars, we manually retrieved light curves using the \texttt{lightkurve} package by selecting the target pixels directly and defining custom photometric apertures. This hybrid approach ensured complete coverage of the original target list.

To ensure uniform frequency sensitivity across all stars, we re-binned the stitched light curves to a 10-minute cadence, matching the time resolution of the majority of sectors. This resulted in a consistent Nyquist frequency of $\sim$72\,d$^{-1}$, suitable for detecting both $\delta$~Sct and $\gamma$~Dor pulsations.

\subsection{Frequency extraction and initial classification}
To characterise the variability and identify pulsational behaviour, we visually inspected the processed light curves and their frequency spectra. The classifications were based on the distribution and morphology of peaks in frequency space. As described in Sect.~\ref{1}, p-mode pulsations generally occur at frequencies above $5\,\mathrm{d}^{-1}$, while g-mode pulsations are found predominantly below $5\,\mathrm{d}^{-1}$. We therefore classified stars showing multiple narrow peaks mainly above this frequency as p-mode pulsators, and those with multiple narrow peaks mainly below it as g-mode pulsators. Rotational variables were identified from low-frequency modulation accompanied by harmonics of the dominant rotational frequency, while eclipsing binaries were recognised from their characteristic light-curve morphology and harmonic structure in the frequency spectrum. From the final sample of 149 stars, 56 stars showed variability. These include nine p-mode pulsators and two g-mode pulsators. In addition, we identified three stars as candidate eclipsing binaries based on their light curves. We classified 34 stars as low-amplitude rotational modulation and eight stars showed irregular variability. The remaining stars showed uncertain variability classifications or no clear variability above the noise level.

To extract mode frequencies for subsequent asteroseismic analysis of the identified p-mode and g-mode pulsators, we applied an iterative pre-whitening procedure using the open source package \texttt{star\_shadow} \citep{IJspeert2024}. In each iteration, the highest-amplitude frequency in the periodogram is fitted with a sinusoidal model and subtracted from the full light curve, after which the residual periodogram is recomputed. This procedure is repeated until no additional significant signal remains. The significance threshold is determined internally by \texttt{star\_shadow} for each star, based on the signal-to-noise ratio of the extracted peaks and the Bayesian Information Criterion (BIC) of the iterative model fit.

\section{Asteroseismology}
\label{4}

\subsection{Pressure mode pulsators}
\subsubsection{Pulsational properties} \label{411}

\begin{figure}
    \centering
    \includegraphics[width=1\linewidth]{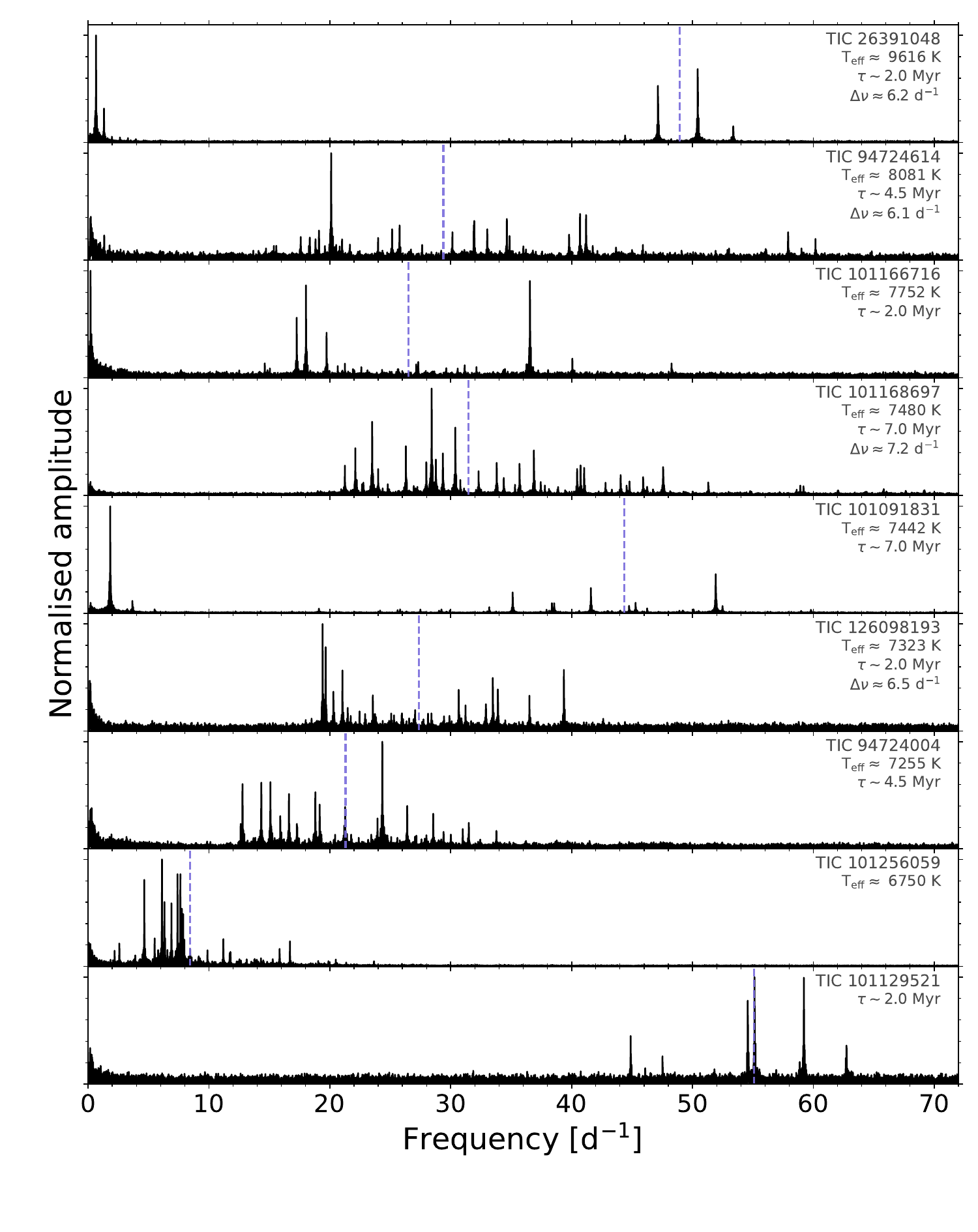}
    \caption{Frequency spectra of the eight $\delta$~Sct stars identified in this work, along with V705~Per (TIC 101256059), ordered by decreasing $T_\mathrm{eff}$ from top to bottom. TIC~101129521, for which no reliable effective-temperature estimate was available, is shown last. The amplitudes are normalised for each star. $\tau$ indicates the adopted ages of the parent subgroups for the eight identified $\delta$~Sct stars.} Approximate values of the $\Delta\nu$ inferred from the échelle diagrams are indicated where available. The dashed vertical lines mark the adopted $\nu_{\mathrm{max}}$ values, calculated as the average of the four estimation methods described in Sect.~\ref{nu_max_methods}.
    \label{dsct}
\end{figure}

In this section, we examine the pulsational properties of the identified $\delta$~Sct stars by analysing their frequency spectra. The frequency spectra of the nine $\delta$~Sct stars considered in this work are shown in Fig.~\ref{dsct}. Eight of these stars were identified among the candidates selected from the \emph{Gaia}-based Perseus membership catalogues, while V705~Per (TIC~101256059) was included because its $\delta$~Sct pulsation has previously been established. The stars with available effective temperature estimates are ordered from top to bottom by decreasing $T_\mathrm{eff}$. For most targets, we adopted the \emph{Gaia} GSP-Phot temperatures. Where these were unavailable, we used spectroscopic temperatures from the literature: for TIC~101166716 and TIC~126098193, we adopted the values from \citet{Sprague_2022}, while for V705~Per we used the temperature reported by \citet{ripepi}. For TIC~101129521, no reliable $T_\mathrm{eff}$ estimate could be found in the literature; it is therefore displayed at the end of the figure.

The eight $\delta$~Sct stars selected from the \emph{Gaia}-based sample exhibit multiple high-frequency peaks predominantly between 20 and 60\,d$^{-1}$, whereas V705~Per extends the observed frequency range to substantially lower values. With the stars arranged by decreasing effective temperature, a general shift towards lower pulsation frequencies is visible from top to bottom of Fig.~\ref{dsct}. In particular, V705~Per is the coolest star with an available $T_\mathrm{eff}$ estimate and also exhibits the lowest dominant pulsation frequency in the sample. The progression is not strictly monotonic, however, and considerable variation remains between individual stars. Even within young populations of similar age and composition, $\delta$~Sct stars can exhibit highly diverse frequency spectra \citep{bedding23}. Differences in stellar mass, evolutionary stage and rotation may introduce additional scatter into the observed temperature--frequency distribution \citep{murphy23,Murphy2024,Gautam2026}.

Inspection of the échelle diagrams revealed approximate regular frequency spacings for four of the $\delta$~Sct stars (see Fig. \ref{echelle}). The corresponding estimates of the large-frequency separation, $\Delta\nu$, are indicated in Fig.~\ref{dsct}. We regard these values as approximate because the ridges are not equally well defined in every target and no formal uncertainties have been derived. The large-frequency separation is related to the acoustic travel time through the star and scales approximately with the square root of its mean density, making it a useful first-order seismic diagnostic \citep{Suarez,murphy23,Gautam2026}. Regular high-frequency patterns have been detected in many young $\delta$~Sct stars, although they are not equally apparent in every object, particularly when the pulsation spectrum is affected by rapid rotation \citep{bedding20,Murphy2024,Gautam2026}. Detailed mode identification and seismic modelling would be required before the approximate spacings measured here could be used to derive precise stellar properties.

The pulsation frequencies can also be considered in the context of evolutionary stage. \citet{Zwintz2014} showed observationally that the highest detected pulsation frequency, $f_{\mathrm{max}}$, of PMS $\delta$~Sct stars generally increases as the stars evolve towards the ZAMS. The models of \citet{Gautam2026} similarly show that radial and non-radial p-mode frequencies generally increase during PMS contraction as the mean stellar density rises, reaching their highest values near the ZAMS before gradually decreasing during the main-sequence phase. Their model grid includes moderate rotation from the beginning of the PMS phase, but adopts standard chemically homogeneous and initially fully convective PMS models rather than following the formation of the stars through protostellar accretion. Among our eight identified $\delta$~Sct stars, six are associated with Perseus subgroups younger than approximately 5~Myr, while the remaining two are associated with Gorgophone, which has an estimated age of approximately 7~Myr \citep{Olivares2023,pavlidou26}. However, the two stars associated with Gorgophone do not show systematically higher $f_{\mathrm{max}}$ values than the stars belonging to the younger subgroups. Rotation can contribute substantially to this scatter, since centrifugal deformation lowers the mean stellar density and hence the acoustic-mode frequencies. At critical rotation, the mode frequencies are reduced to approximately two-thirds of their non-rotating values, while for rotation rates typical of A-type stars the reduction is still of order 10\% \citep{murphy2026}. The observed $f_{\mathrm{max}}$ values therefore remain scattered across the sample, likely reflecting differences in stellar mass and rotation, as well as uncertainties in the subgroup and individual stellar ages.

V705~Per (TIC~101256059), also known as H254, was first identified as a PMS $\delta$~Sct star in IC~348 by \citet{ripepi}. The authors confirmed that its dominant frequency near 7.4\,d$^{-1}$ corresponds to a radial pulsation mode and identified it as the fundamental radial mode. From its position in the HR diagram, they estimated a mass of approximately $2.2\,M_\odot$ and an age of $5\pm1$~Myr. On the basis of this established pulsational classification, we include V705~Per among the $\delta$~Sct stars analysed here.

Its membership in IC~348 is less secure than its pulsational classification. V705~Per is not included in the \emph{Gaia}-based membership lists of \citet{Olivares2023} or \citet{Pavlidou2021}. In addition, \citet{zhou} identified it as a proper-motion outlier and noted that its status as a possible walk-away star from IC~348 remains uncertain. However, we treat V705~Per as a pulsating cluster member for the purpose of the following analysis.
The position of V705~Per at the cool and low-frequency end of Fig.~\ref{dsct} is consistent with its previously identified fundamental radial mode. The models of \citet{Gautam2026} show that the fundamental radial mode and other low-order p modes can occur at relatively low frequencies during the PMS phase, with their frequencies increasing as the star contracts towards the ZAMS. V705~Per therefore extends the qualitative temperature--frequency progression visible across the sample and its frequency spectrum is compatible with the interpretation of a young PMS $\delta$~Sct star.

The same models show that low-order g modes and f modes can occur close to the expected frequency of the fundamental radial mode. Such modes could contribute to some of the additional peaks visible in the spectrum of V705~Per, although mode identification would be required to establish their nature. The pulsation spectrum supports the established $\delta$~Sct classification of V705~Per and is consistent with a young evolutionary stage, but it cannot independently establish its age or membership in IC~348. Yet, evolved main sequence stars near the terminal-age main sequence exhibit similar pulsation frequencies \cite{Gautam2026} and only detailed modelling can distinguish between their evolutionary phases (Tedeskino et al., in prep.).

\subsubsection{Estimating $\nu_{\mathrm{max}}$} \label{nu_max_methods}
To characterise the overall pulsation properties of each identified $\delta$~Sct star, we estimated $\nu_{\mathrm{max}}$, following the methodology described by \citet{Murphy2024}. The parameter $\nu_{\mathrm{max}}$ denotes the frequency at which the pulsation amplitude or power is most strongly concentrated in the observed spectrum. It provides a global description of the pulsation content and enables comparison between stars even when individual mode identification is not possible.

To minimise method-dependent biases, $\nu_{\mathrm{max}}$ was computed using four independent approaches based on the set of statistically significant frequencies extracted from the pre-whitening procedure.

\begin{enumerate}
    \item {Amplitude-weighted mean frequency}:
    \[
    \nu_{\rm amp} = \frac{\sum_i A_i \nu_i}{\sum_i A_i},
    \]
 which emphasises frequencies with the largest observed amplitudes.

    \item {Power-weighted mean frequency}:
    \[
    \nu_{\rm power} = \frac{\sum_i A_i^2 \nu_i}{\sum_i A_i^2},
    \]
    following \citet{barcelo}, which gives additional weight to the highest-power modes.

    \item {Amplitude smoothing}: A Gaussian kernel was applied to the amplitude spectrum, and $\nu_{\mathrm{max}}$ was taken as the frequency at maximum smoothed amplitude.

    \item {Power smoothing}: The same procedure was applied to the power spectrum, and $\nu_{\mathrm{max}}$ was defined as the frequency at maximum smoothed power.
\end{enumerate}


For the smoothing methods, we adopted a fixed Gaussian kernel width of $30\,\mathrm{d}^{-1}$, representative for young $\delta$~Sct stars ($\sim\!4\Delta\nu$) \citep{bolouki2026,murphy2026}, following the methodology applied by \citet{Murphy2024}. Using a fixed kernel width allowed the same smoothing procedure to be applied consistently across the full sample.

The final $\nu_{\mathrm{max}}$ value for each star was taken as the mean of the four estimates. The four methods yield estimates within approximately \(3\,\mathrm{d}^{-1}\) of one another for most stars, although a larger spread is found for V705~Per (see Fig.~\ref{app1}). This indicates that the determination of $\nu_{\mathrm{max}}$ is robust against the choice of method and suggests that the pulsation power is distributed over a well-defined frequency range rather than being dominated by a single isolated mode.

\subsubsection{The $\nu_{\mathrm{max}}$--$T_\mathrm{eff}$ relation}

To place our results in a broader asteroseismic context, we examine the relation between the frequency of maximum pulsation power, $\nu_{\mathrm{max}}$, and effective temperature. This relation provides a global characterisation of the pulsation spectra of $\delta$~Sct stars and has been explored as a potential empirical scaling relation \citep{Murphy2024} analogous to that observed in solar-like oscillators, where $\nu_{\mathrm{max}}$ is closely linked to fundamental stellar parameters \citep{brown1991,kjeldsen1995}.

In $\delta$~Sct stars, however, previous studies have shown that any such relation is considerably less well defined. While correlations between $\nu_{\mathrm{max}}$ and $T_{\mathrm{eff}}$ have been reported \citep{barcelo,bowman,Murphy2024}, they are typically characterised by substantial scatter. In particular, \citet{Murphy2024} analysed a large sample of young main-sequence $\delta$~Sct stars in the Cep--Her Complex and derived an empirical scaling relation between $\nu_{\mathrm{max}}$ and $T_{\mathrm{eff}}$. However, they found that the scatter in this relation is large and that the observed distribution is not well described by a single continuous sequence.

Instead, \citet{Murphy2024} showed that the $\nu_{\mathrm{max}}$--$T_{\mathrm{eff}}$ diagram exhibits two distinct ridges. One ridge broadly follows the expected trend in which $\nu_{\mathrm{max}}$ increases with effective temperature, while the second ridge is offset toward lower $\nu_{\mathrm{max}}$ values at a given $T_{\mathrm{eff}}$. This structure indicates that stars with similar temperatures can exhibit significantly different dominant pulsation frequencies.

The origin of this two-ridge structure was interpreted in terms of stellar rotation. \citet{Murphy2024} argued that rapidly rotating stars preferentially excite low-order pressure modes, which concentrate pulsation power at lower frequencies compared to the higher-order modes typically observed in more slowly rotating stars. As a result, rapid rotators tend to populate the ridge at lower $\nu_{\mathrm{max}}$. In addition, rotation modifies the stellar structure through centrifugal deformation and changes in mean density, further contributing to the observed spread in $\nu_{\mathrm{max}}$.

\begin{figure}
    \centering
    \includegraphics[width=1\linewidth]{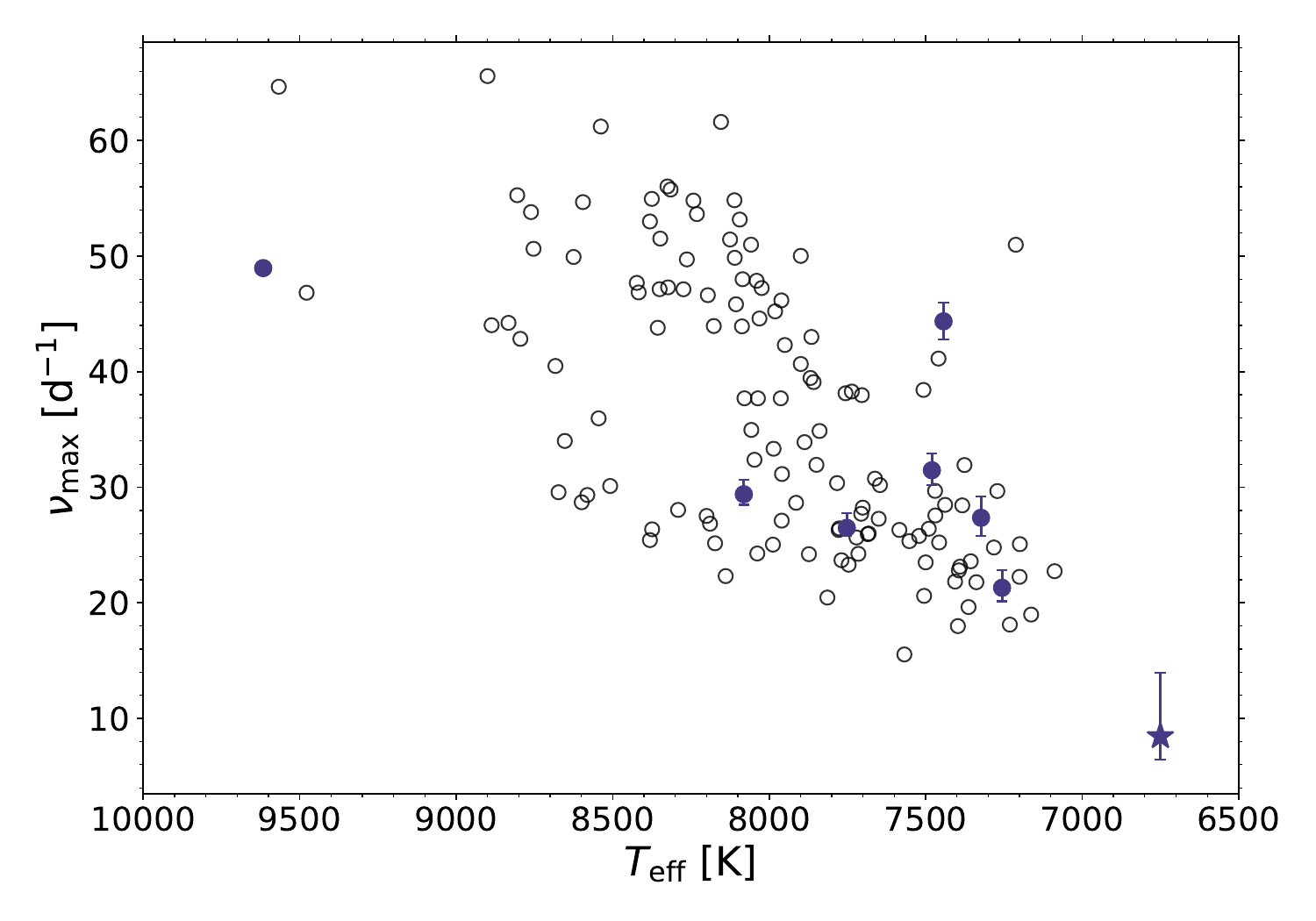}
    \caption{$\nu_{\mathrm{max}}$ against the effective temperature $T_\mathrm{eff}$ for $\delta$~Sct stars. Purple points show the young $\delta$~Sct stars identified in this work along with the previously identified $\delta$~Sct star V~705 Per as a star symbol, while the open circles represent the main-sequence sample from \citet{Murphy2024}. Error bars on the purple points indicate the range of $\nu_{\mathrm{max}}$ values obtained using four different estimation methods. Our stars occupy a similar range in $\nu_{\mathrm{max}}$ as the comparison sample, and show a decrease in $\nu_{\mathrm{max}}$ with $T_\mathrm{eff}$.}
    \label{fig:teff}
\end{figure}

Figure~\ref{fig:teff} shows $\nu_{\mathrm{max}}$ as a function of $T_{\mathrm{eff}}$ for the $\delta$~Sct stars in our sample, together with the Cep--Her sample from \citet{Murphy2024}. Most of our stars occupy a relatively narrow temperature range around $T_{\mathrm{eff}} \sim 7200$--$8000\,\mathrm{K}$ and exhibit a significant spread in $\nu_{\mathrm{max}}$, spanning approximately $20$ to $55\,\mathrm{d}^{-1}$. Within this limited temperature range, no clear internal trend is apparent, and stars with similar $T_{\mathrm{eff}}$ display markedly different $\nu_{\mathrm{max}}$ values.

When compared to the Cep--Her sample, our targets fall within the same overall parameter space and extend the observed distribution toward younger evolutionary stages. In particular, the range of $\nu_{\mathrm{max}}$ values observed in our sample overlaps with both ridges identified by \citet{Murphy2024}, although the small number of stars and their temperature coverage prevents a clear separation into distinct sequences.

A direct test of the rotation-driven interpretation is also not possible for our sample. The \emph{Gaia} \texttt{vbroad} parameter, a proxy for rotational broadening, is not available for most of our targets.

Nevertheless, the presence of substantial scatter in $\nu_{\mathrm{max}}$ at similar effective temperatures, together with the overall agreement with the Cep--Her distribution, suggests that rotation may play a comparable role in shaping the pulsation properties of very young $\delta$~Sct stars. Establishing whether the two-ridge structure persists into the pre-main sequence regime will require a larger sample with well-constrained stellar parameters and independent measurements of rotation.

\subsection{Gravity mode pulsators}
\begin{figure}[htbp]
    \centering
    \includegraphics[width=1\linewidth]{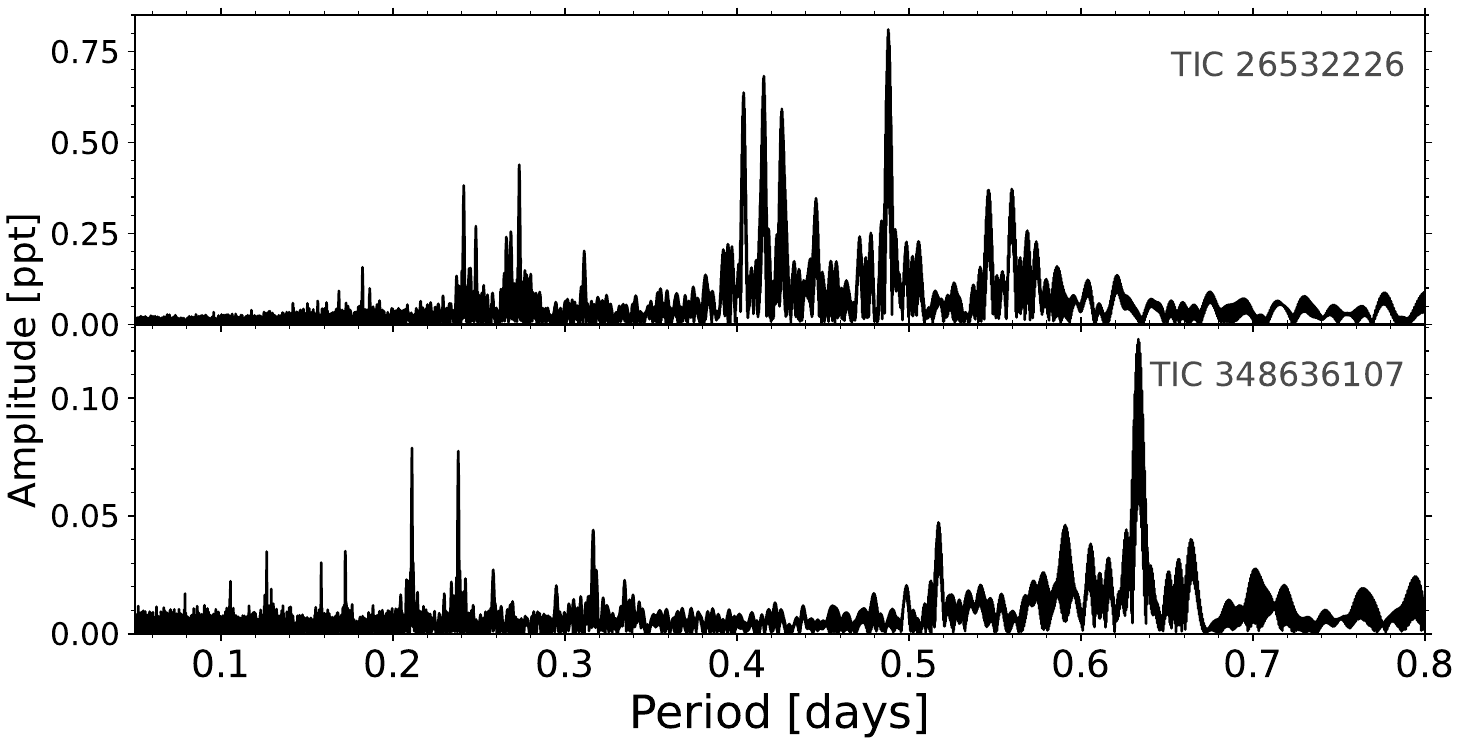}
    \caption{Periodograms of HD~281229 and HD~23244, the two gravity mode pulsators in our sample. HD~281229 shows a clear g-mode period-spacing pattern, while the classification of HD~23244 remains less certain due to loosely spaced peaks and no clear pattern.}
    \label{gdor_pg}
\end{figure}

Having discussed the p-mode pulsators, we now look at the stars whose variability is concentrated in the low-frequency regime. Two targets stand out in this part of the frequency spectrum: HD~281229 (TIC~26532226) and HD~23244 (TIC~348636107). Both show groups of closely spaced peaks at periods below approximately 1\,d (Fig.~\ref{gdor_pg}).

HD~281229 has an effective temperature of approximately \(7350\,\mathrm{K}\), placing it within the expected temperature range of $\gamma$~Dor stars. Its periodogram shows two concentrations of peaks, one between  0.2\,d and 0.3\,d and another between 0.4\,d and 0.5\,d. Within these regions, a sequence of nearly regularly spaced peaks is visible (Fig.~\ref{gdor_pg}). This structure is indicative of a high-order g-mode period-spacing pattern and allows the star to be classified as a $\gamma$~Dor star.

HD~23244 exhibits dominant variability at somewhat longer periods, mainly between approximately 0.6 and 0.7\,d. However, no clear or continuous period-spacing pattern can be identified. At these longer periods, the limited TESS observing baseline reduces the period resolution, making it more difficult to resolve neighbouring pulsation modes and identify a regular spacing pattern. HD~23244 is a hotter and more luminous B7 star, so its low-frequency variability could be caused by high-order g modes characteristic of a young SPB star. However, in the absence of a clear period-spacing pattern and lesser significant peaks, rotational modulation cannot be ruled out. We therefore regard HD~23244 as a candidate SPB star or rotational variable.

\subsubsection{Fitting the period-spacing pattern of HD~281229}

\begin{figure*}[!t]
    \centering
    \includegraphics[width=\textwidth]{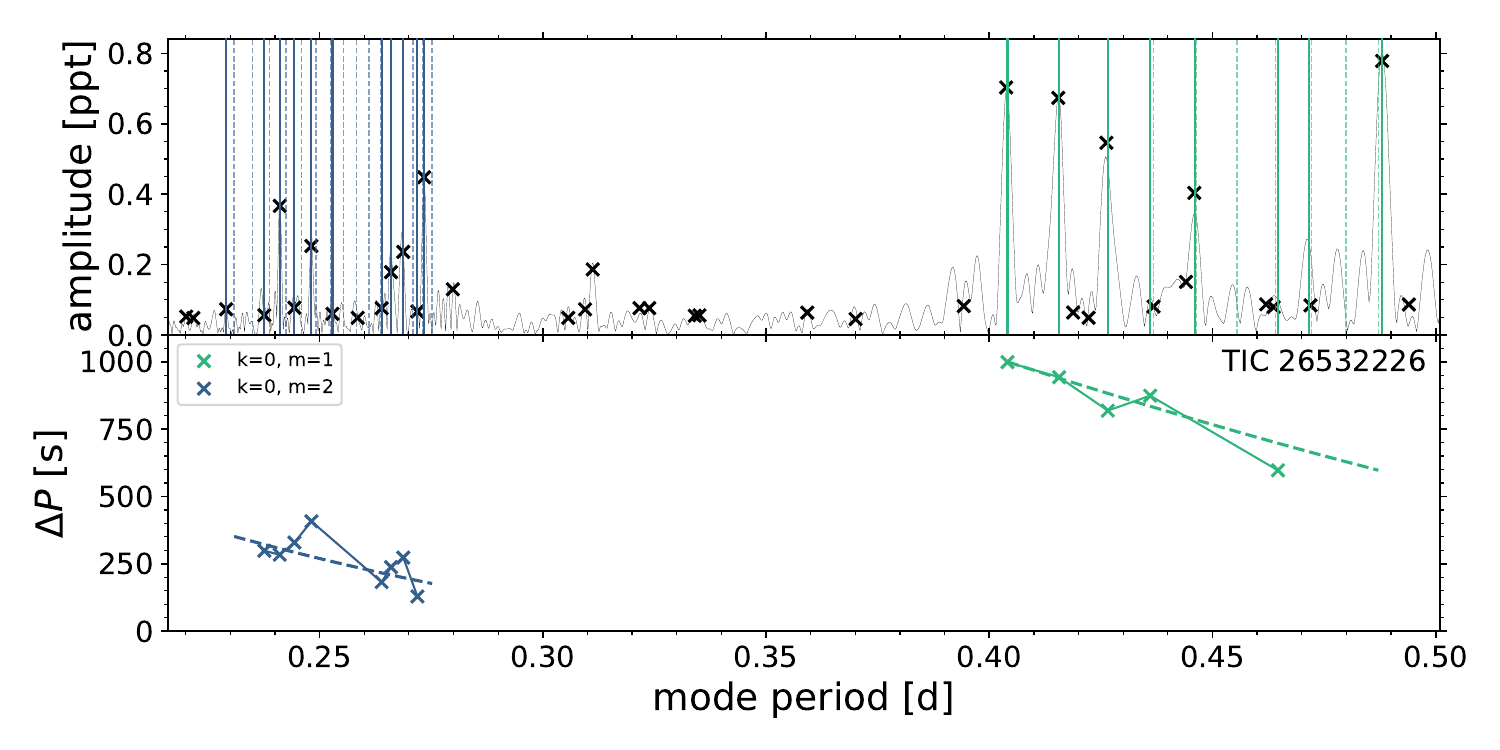}
    \caption{Period-spacing pattern for prograde quadrupole (blue) and dipole (green) modes of HD~281229. The top panel shows the amplitude spectrum in period space. The cross symbols denote the pre-whitened frequencies, while the solid vertical lines mark the selected modes of the period-spacing pattern. The dashed vertical lines indicate the best-fit asymptotic pulsation solution. The bottom panel shows the corresponding period spacings, $\Delta P$, as a function of period, together with the best asymptotic fit (dashed).}
    \label{sim_psp}
\end{figure*}

To analyse the period-spacing patterns of HD~281229, we used the \texttt{AMiGO} software package \citep{VanReeth2016}. This tool fits observed period-spacing patterns with analytical expressions derived under the Traditional Approximation of Rotation \citep[TAR,][]{townsend,mathis} by minimising the $\chi^2$ difference between the observed and theoretical spacings. Two distinct patterns were identified and fitted simultaneously: a prograde dipole sequence with $(l,m)=(1,1)$ and a prograde quadrupole sequence with $(l,m)=(2,2)$. The resulting period-spacing patterns and their best-fitting models are shown in Fig.~\ref{sim_psp}. The fit yields an asymptotic period spacing of $\Pi_0 = 4776 \pm 400$\,s and a near-core rotation frequency of $f_{\mathrm{rot}} = 1.317 \pm 0.05$\,d$^{-1}$. 

    


\subsubsection{Evolutionary context of $\Pi_0$}

The key asteroseismic quantity $\Pi_0$ captures information about the near-core structure in a single parameter. In the asymptotic regime, $\Pi_0$ is related to the integral of the Brunt--V\"ais\"al\"a frequency across the g-mode propagation cavity, and therefore reflects the stratification of the deep stellar interior \citep{garcia,Aerts2010}. Deviations from an asymptotic period-spacing pattern arise when structural features near the core, such as chemical gradients and mixing at the convective core boundary, modify the local buoyancy structure and introduce characteristic dips or oscillatory signatures in the pattern \citep{miglio2008,Pedersen2018,Michielsen2021,mombarg21}.

\begin{figure}[!htpb]
    \centering
    \includegraphics[width=1\linewidth]{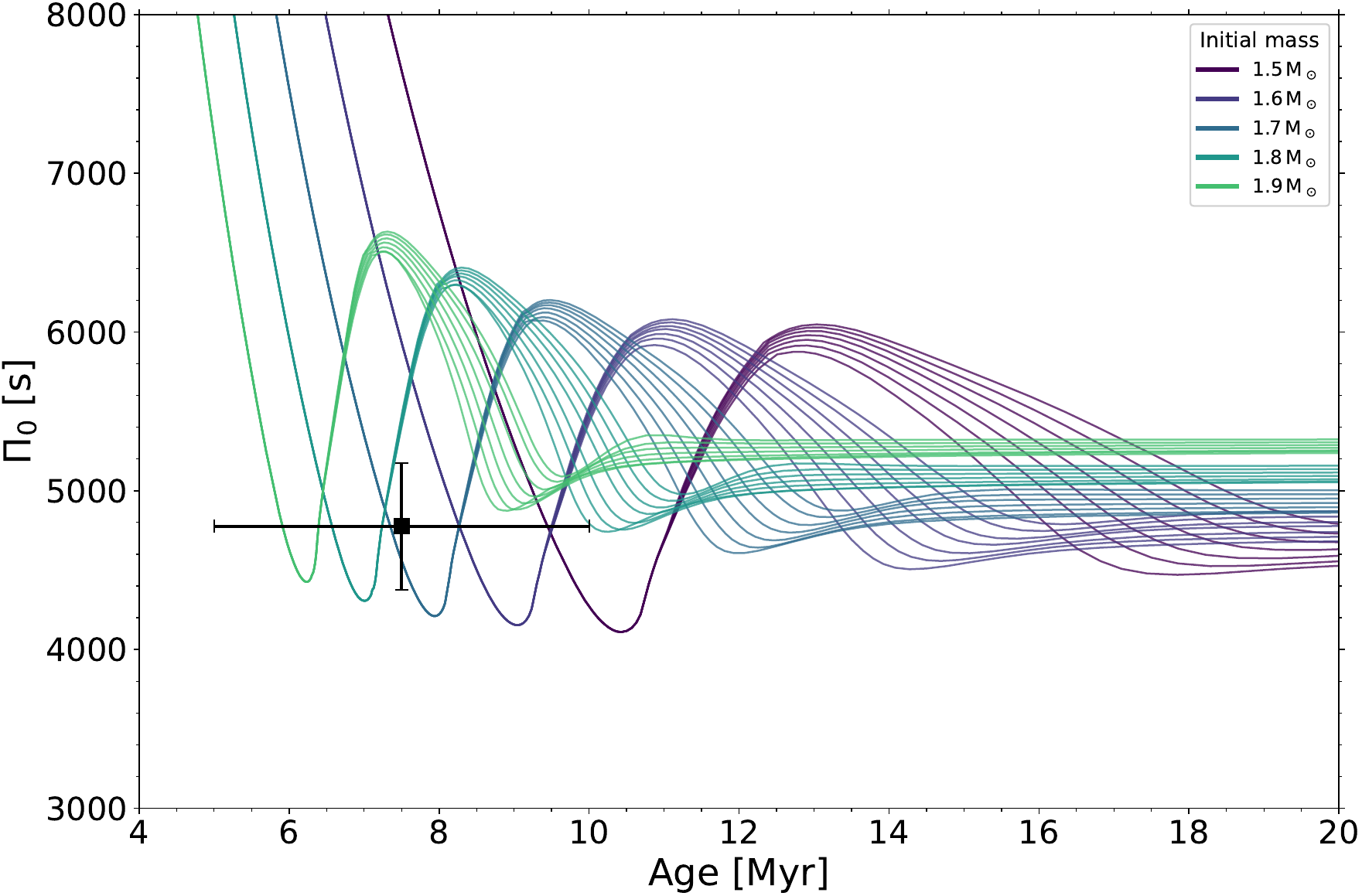}
    \caption{Asymptotic period spacing, $\Pi_0$, as a function of age for MESA evolutionary tracks from \citet{mombarg24}, with initial masses between $1.5$ and $1.9\,\mathrm{M}_\odot$. For each combination of mass and convective-core overshooting parameter, we retain the rotational sequence whose near-core rotation frequency at the ZAMS is closest to the value measured for HD~281229, $f_{\mathrm{rot}}=1.317\,\mathrm{d}^{-1}$. The seven tracks shown at each mass correspond to overshooting parameters ranging from $f_{\mathrm{ov}}=0.005$ to $0.035$. The black square marks the measured $\Pi_0$ of HD~281229, with the vertical error bar indicating its uncertainty and the horizontal error bar representing the estimated age range of its parent cluster, Alcaeus. The overlap between the observed $\Pi_0$ and the evolutionary tracks is consistent with HD~281229 being a young intermediate-mass star near the ZAMS transition.}
    \label{pi0}
\end{figure}

To interpret the derived value of $\Pi_0$ in an evolutionary framework, we compared the value obtained with a grid of MESA evolutionary tracks that describe the evolution of $\Pi_0$ as a function of age and stellar mass \citep[Fig.~\ref{pi0},][]{mombarg24}. Within the estimated age range of the Alcaeus subgroup, the uncertainty region of HD~281229 overlaps the evolutionary track corresponding to an initial mass of approximately $1.5\,M_\odot$. This is in good agreement with the independent mass estimate of $1.52^{+0.07}_{-0.10}\,M_\odot$ reported by \citet{hunt2024}, based on PARSEC isochrone modelling, and provides a useful consistency check between the seismic and isochrone-based constraints.

This comparison should be treated with some caution. The MESA tracks describe standard PMS contraction and do not include an accretion-informed evolutionary history, while rotation was also not included during the PMS phase. At present, models that combine a realistic treatment of PMS accretion with predictions for the evolution of $\Pi_0$ are not available, hence, the comparison remains approximate. Even so, the position of HD~281229 in the $\Pi_0$--age diagram shows that $\Pi_0$ can provide an additional constraint on the early evolutionary state of young intermediate-mass stars.

\subsubsection{Near-core rotation estimates}
The near-core rotation frequency, $f_{\mathrm{rot}}$, probes the rotation of the deep radiative interior sampled by g modes. Along with the value obtained by the fit for HD~281229, we estimated $f_{\mathrm{rot}}$ for both g-mode pulsators using the regression fit of \citet{aerts2025evolutionnearcorerotationfrequency} (Eq.~4) that quantifies the observational relation between $f_\mathrm{rot}$ and the frequency of the dominant mode, $f_\mathrm{dom}$. This relation was calibrated using 105 Kepler $\gamma$~Dor and SPB stars whose dominant mode was identified as a prograde dipole gravito-inertial mode in the sub-inertial regime and had an observed amplitude above 4\,mmag.


The uncertainties quoted for the slope and intercept describe the uncertainty in the fitted regression coefficients and do not represent the prediction uncertainty for an individual star. From a bootstrap analysis of the regression residuals, \citet{aerts2025evolutionnearcorerotationfrequency} obtained an asymmetric $1\sigma$ prediction interval of $-0.164$ and $+0.123\,\mathrm{d}^{-1}$. We therefore adopt this empirical interval for the rotation frequencies inferred from $f_{\mathrm{dom}}$.

Applying the relation to the dominant mode associated with the period-spacing pattern of HD~281229 gives
$f_{\mathrm{rot}}=1.441^{+0.123}_{-0.164}\,\mathrm{d}^{-1}.$
For HD~23244, we obtain $f_{\mathrm{rot}}=1.047^{+0.123}_{-0.164}\,\mathrm{d}^{-1}.$

For HD~281229, the empirical estimate is consistent within the quoted uncertainties with the value obtained from the \texttt{AMiGO} period-spacing fit, although its central value is higher. We adopt the \texttt{AMiGO} result as the preferred estimate because it uses both detected period-spacing patterns, whereas the empirical relation relies on the frequency of a single dominant mode in the dipole sequence. The latter method is known to exhibit greater scatter and stronger systematic deviations than rotation measurements based on period-spacing patterns \citep{aerts2025evolutionnearcorerotationfrequency,li26}.

\subsection{Stars with rotational modulation}

\begin{figure}[!htpb]
    \centering
    \includegraphics[width=1\linewidth]{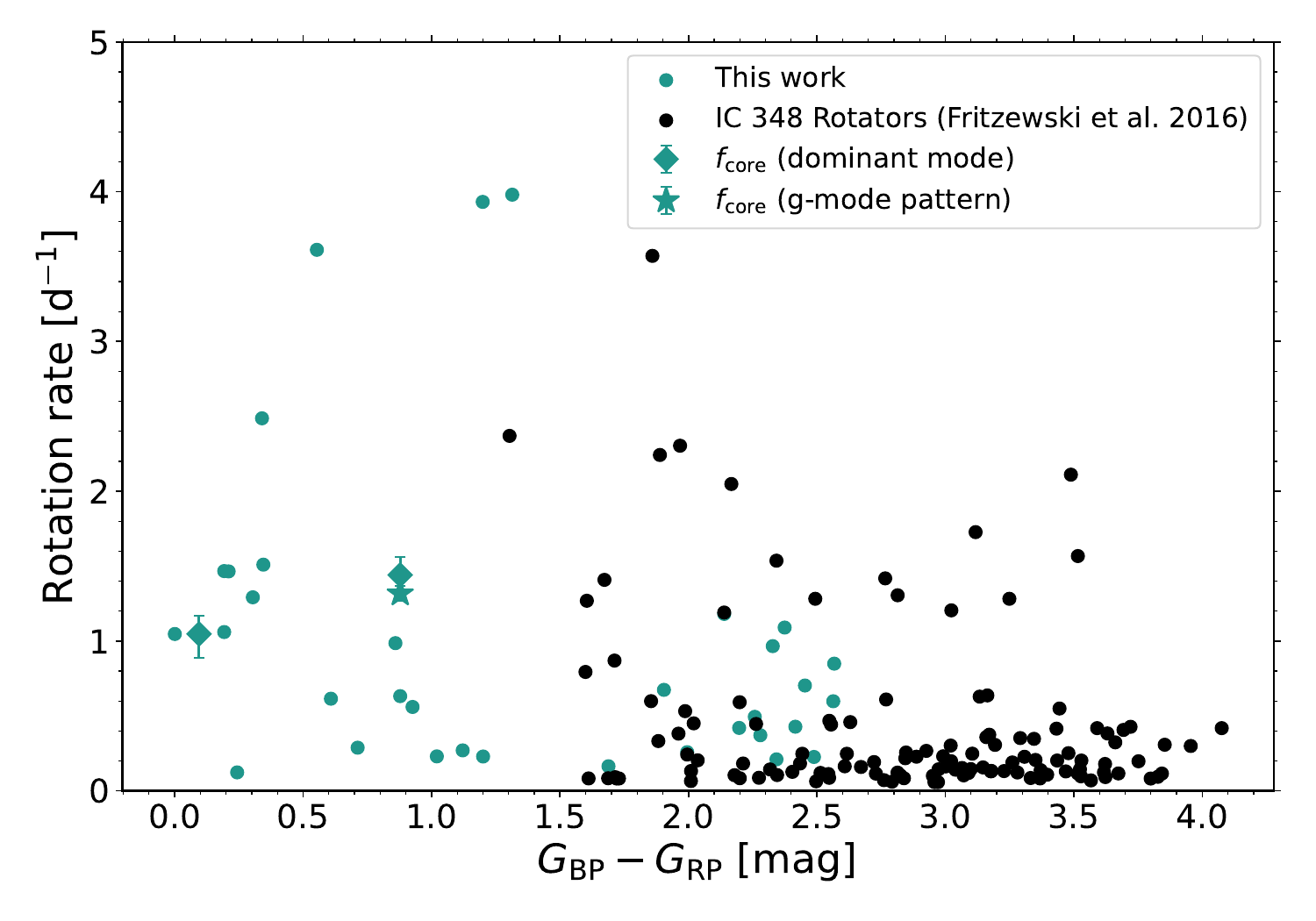}
    \caption{Rotation frequency as a function of \emph{Gaia} colour $G_{\mathrm{BP}}-G_{\mathrm{RP}}$. Teal points show the rotators identified in this work, while black circles show the IC~348 rotators from \citet{Fritzewski_2016}. The diamonds and star indicate the near-core rotation frequencies of the g-mode candidate pulsators derived from the dominant mode and the g-mode period-spacing pattern, respectively.}
    \label{rot}
\end{figure}

In addition to the pulsating stars, we identified 34 stars in our sample whose variability is dominated by rotational modulation. We determined their rotation rates and compared them with previously measured rotators in IC~348. Figure~\ref{rot} shows the rotation rates as a function of \emph{Gaia} colour, $G_{\mathrm{BP}}-G_{\mathrm{RP}}$, together with the IC~348 sample of \citet{Fritzewski_2016}. The two samples overlap at intermediate colours, but our sample extends the available rotation measurements substantially towards bluer colours, corresponding to more massive stars. This provides rotation measurements for a part of the young stellar population that is only sparsely represented in the previous IC~348 sample. 

For comparison, we show the near-core rotation rates of the two g-mode pulsators. They fall well within the observed range of the surface-modulated stars.

Across the bluer part of the combined distribution, the rotation rate tends to decrease towards redder colours. Albeit, we also find some very fast rotating stars at 4\,d$^{-1}$. Beyond $G_{\mathrm{BP}}-G_{\mathrm{RP}}\approx 2.0$, the distribution shows more slowly rotating stars ($f_\mathrm{rot}\lesssim0.5$\,d$^{-1}$). Yet, we also find stars rotating significantly faster. This is expected for such a young PMS population, where a well-defined colour--rotation sequence has not yet fully developed \citep{rebull2020}. The Perseus measurements therefore complement the existing IC~348 rotation sample by extending the observed distribution to higher-mass stars.

The rotation frequencies for the warmer stars measured in Perseus largely overlap with the rotation frequency distributions reported for NGC 2516 and the Pleiades \citep{li24,Fritzewski_2026}. A few Perseus stars reach values close to $4\,\mathrm{d}^{-1}$, which is not observed in the older clusters. In contrast to our sample, the rotation rates of early-type stars in the older clusters are mostly asteroseismic, while we find many surface-modulated stars. As such, we cannot distinguish whether the difference is evolutionary (i.e. stars spinning down) or observational (i.e. surface modulation not observed at older ages). Future observations of g-mode pulsators in young open clusters and associations will eventually help us to understand the rotational evolution of young early-type pulsators.


\section{Conclusions}\label{5}

We report the discovery of a gravity-mode period-spacing pattern in HD~281229 and its modelling with the TAR, which yields constraints on the asymptotic period spacing and near-core rotation. This result highlights the asteroseismic potential of very young stars in the Perseus Molecular Cloud and provides a first detailed probe of their deep internal structure.

More broadly, we carried out a systematic search for pulsating stars near the ZAMS in the Perseus Molecular Cloud using light curves extracted from TESS FFIs. We identified ten pulsators, comprising eight $\delta$~Sct stars and two g-mode candidates. Together with the previously known $\delta$~Sct star V705~Per, these form the sample of eleven pulsators analysed in this work. We also identified 34 stars with rotational modulation.

Frequencies were extracted through iterative pre-whitening and $\nu_{\mathrm{max}}$ was estimated for all p-mode pulsators. The resulting $\nu_{\mathrm{max}}$ values fall within the range reported for young main-sequence $\delta$~Sct stars \citep{Murphy2024}. However, given the modest size of our sample, we cannot confirm any patterns. A larger sample of very young pulsators will be required to determine whether similar relations hold during earlier evolutionary stages.

Among the gravity-mode pulsators, HD~281229 is distinguished by a clear period-spacing pattern characteristic of $\gamma$~Dor stars. By modelling this pattern with \texttt{AMiGO}, we derived its asymptotic period spacing, $\Pi_0$, and near-core rotation rate. To our knowledge, this is the first pre-main sequence and therefore youngest $\gamma$~Dor star for which a period spacing pattern has been modelled and an asymptotic period spacing value measured. This result demonstrates that g-mode asteroseismology can already provide direct constraints on the internal structure and rotation of stars in the PMS phase. 

The 34 stars showing rotational modulation provide an additional view of the young stellar population in the PMC. As expected for a very young population, no well-defined rotational trend is apparent. Nevertheless, the distribution suggests that the bluer stars tend to rotate faster than the redder ones. These measurements provide useful constraints on the rotation properties of the youngest stars in the region.

Significant progress is expected from the ESA PLATO mission. PLATO will obtain long-duration (at least $\sim$2 years per field), high-precision photometric observations using 26 cameras \citep{2025Rauer}. PLATO will observe long-duration fields close to the Galactic plane, where young stellar populations are abundant \citep{nascimbeni}. This increases the likelihood of detecting larger samples of very young pulsators and extending asteroseismic studies toward earlier evolutionary stages. Additionally, the extended baselines will significantly improve the frequency resolution compared to single-sector \textit{TESS} observations, allowing closely spaced gravity modes and rotational splittings to be resolved more reliably. Long and nearly uninterrupted space-based time series have previously enabled detailed studies of g-mode pulsators and the detection of period-spacing patterns \citep{vanreeth2015,Li_2019}. These diagnostics provide direct constraints on near-core rotation and internal mixing processes. In the context of PMS stars, similarly long time series will be essential for robust identification of period-spacing patterns and for placing stronger constraints on the early evolution of the asteroseismic observables $\Pi_0$ and near-core rotation.

With continued TESS observations, upcoming PLATO data, and improved theoretical modelling, pre-main sequence asteroseismology will move from individual detections towards systematic characterisation of young stellar populations. This will provide increasingly stringent constraints on the early structural and rotational evolution of intermediate-mass stars.

\section*{Acknowledgements}

The authors are grateful to Konstanze Zwintz for valuable discussions and feedback. SW thanks the members of the KU Leuven Asteroseismology Research Group for constructive discussions and support throughout this thesis project. 

DJF acknowledges support from the Flemish Government under the long-term structural Methusalem funding program by means of the project SOUL: Stellar evolution in full glory, grant METH/24/012 at KU Leuven. GL acknowledges the support of the Australian Research Council through the DECRA project DE250100773. SJM was supported by the Australian Research Council through Future Fellowship FT210100485. 

This research is based on data collected by the TESS mission, which are publicly available from the Mikulski Archive for Space Telescopes (MAST). Funding for the TESS mission is provided by NASA’s Science Mission Directorate.

This work has made use of data from the European Space Agency (ESA) mission \emph{Gaia} (https://www.cosmos.esa.int/gaia), processed by the \emph{Gaia} Data Processing and Analysis Consortium (DPAC, https://www.cosmos.esa.int/web/gaia/dpac/consortium). Funding for the DPAC has been provided by national institutions participating in the \emph{Gaia} Multilateral Agreement.

This research has made use of NASA’s Astrophysics Data System Bibliographic Services and of the SIMBAD database and the VizieR catalogue access tool, operated at CDS, Strasbourg, France.

\noindent\textbf{Software:} This research made use of \texttt{TOPCAT} \citep{Taylor2005}, \texttt{Astropy} (version 6.0.0; \citealt{Astropy2013}), \texttt{Lightkurve} (version 2.5.0; \citealt{Lightkurve2018}), \texttt{NumPy} (version 1.26.4; \citealt{Harris2020}), \texttt{SciPy} (version 1.12.0; \citealt{Virtanen2020}), \texttt{Matplotlib} (version 3.8.3; \citealt{Hunter2007}), \texttt{pandas} (version 2.2.1; \citealt{McKinney2010}), and \texttt{astroquery} (version 0.4.7; \citealt{Ginsburg2019}).

\bibliographystyle{aa} 
\bibliography{references} 
\clearpage
\appendix
\section{Additional figures and tables}
\begin{center}
\begin{minipage}{\textwidth}
    \centering
    \includegraphics[width=\textwidth]{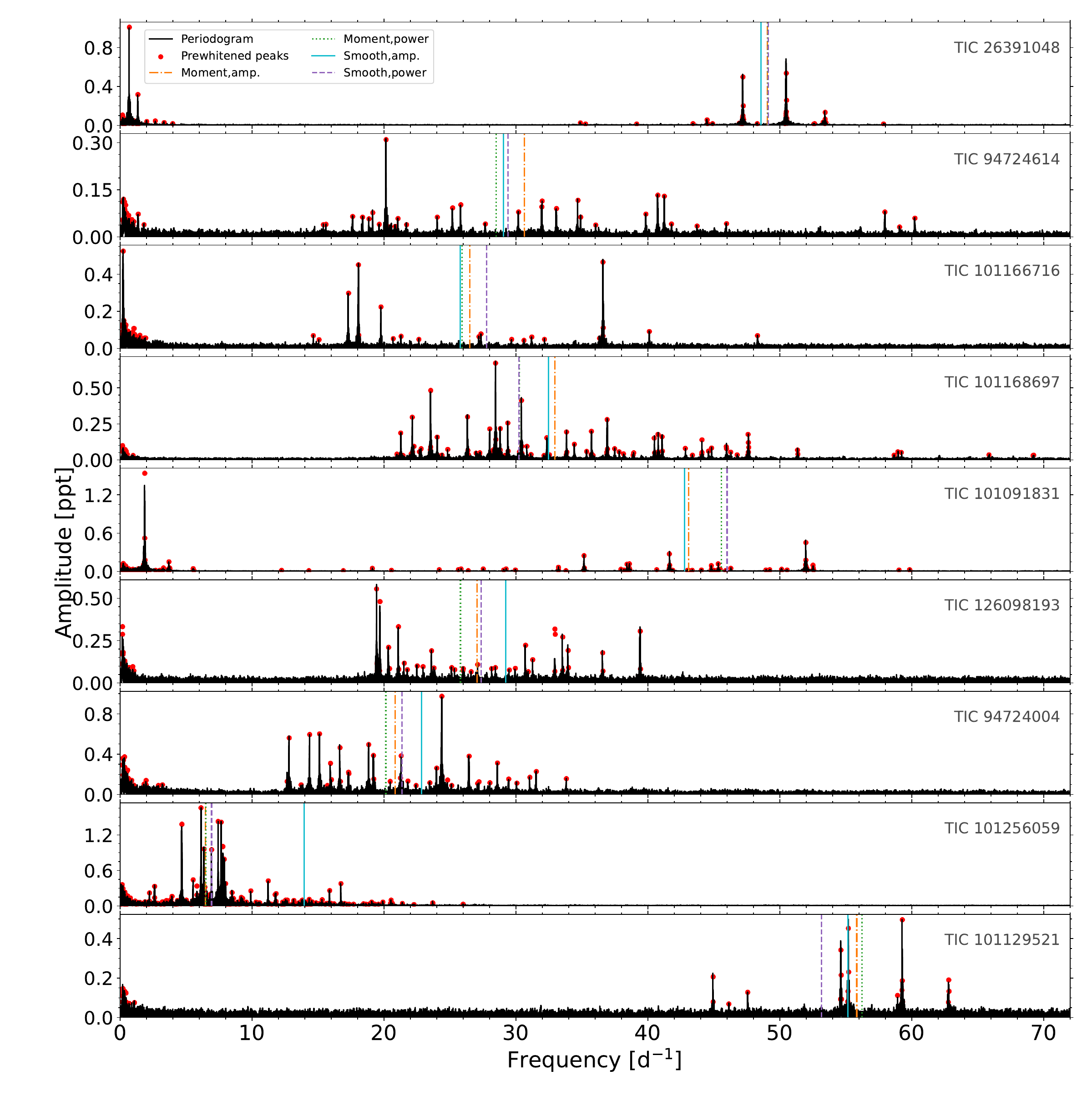}
    \captionof{figure}{Periodograms of the eight $\delta$~Sct stars in our sample, along with V705~Per (TIC~101256059) plotted in frequency space. Coloured vertical lines mark the estimated $\nu_{\mathrm{max}}$ values obtained using four different methods. The red dots represent all the pre-whitened frequencies extracted with \texttt{star\_shadow}.}
    \label{app1}
\end{minipage}
\end{center}

\clearpage

\begin{figure*}
    \centering

    \begin{subfigure}{0.48\textwidth}
        \centering
        \includegraphics[width=\linewidth]{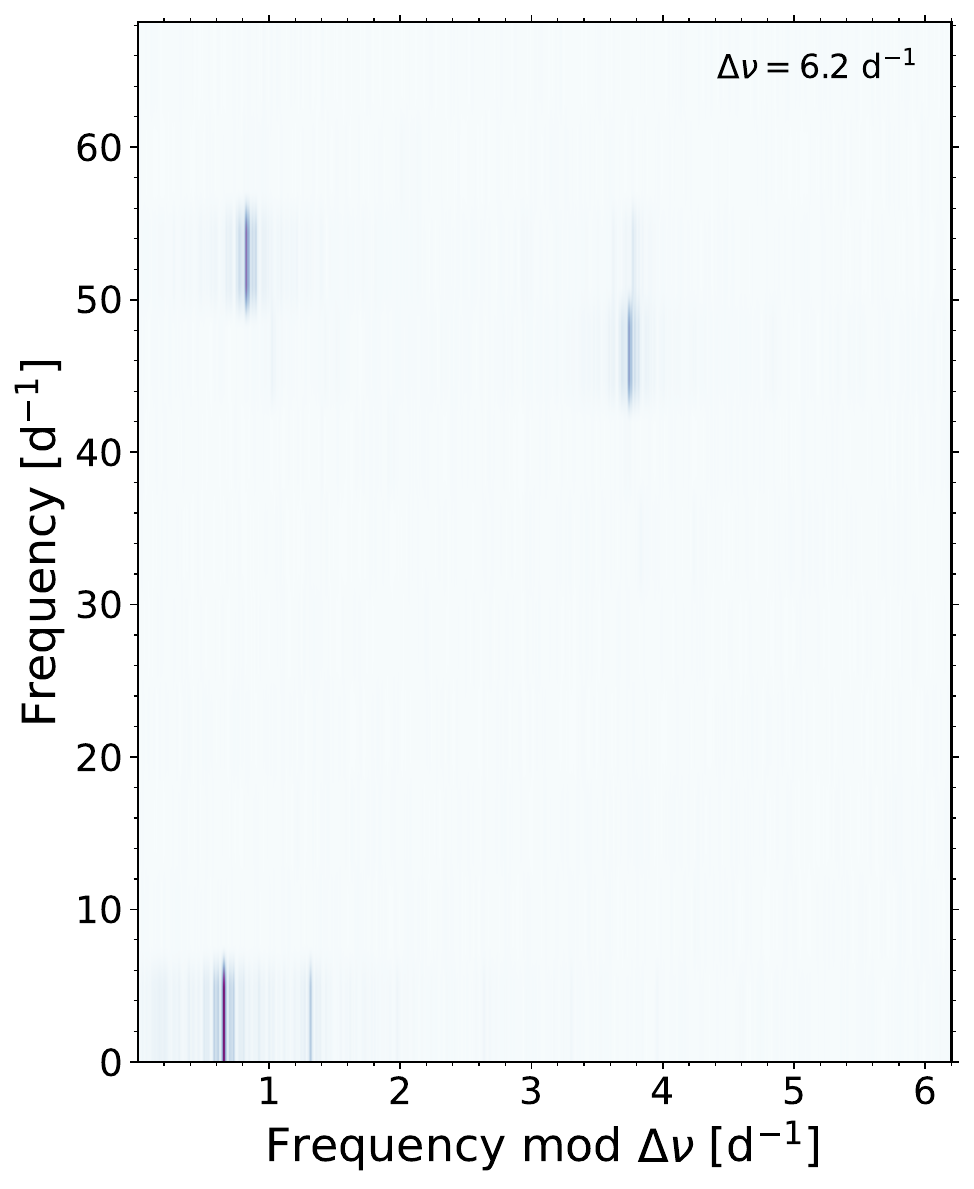}
        \caption{TIC~26391048}
        \label{echelle1}
    \end{subfigure}
    \hfill
    \begin{subfigure}{0.48\textwidth}
        \centering
        \includegraphics[width=\linewidth]{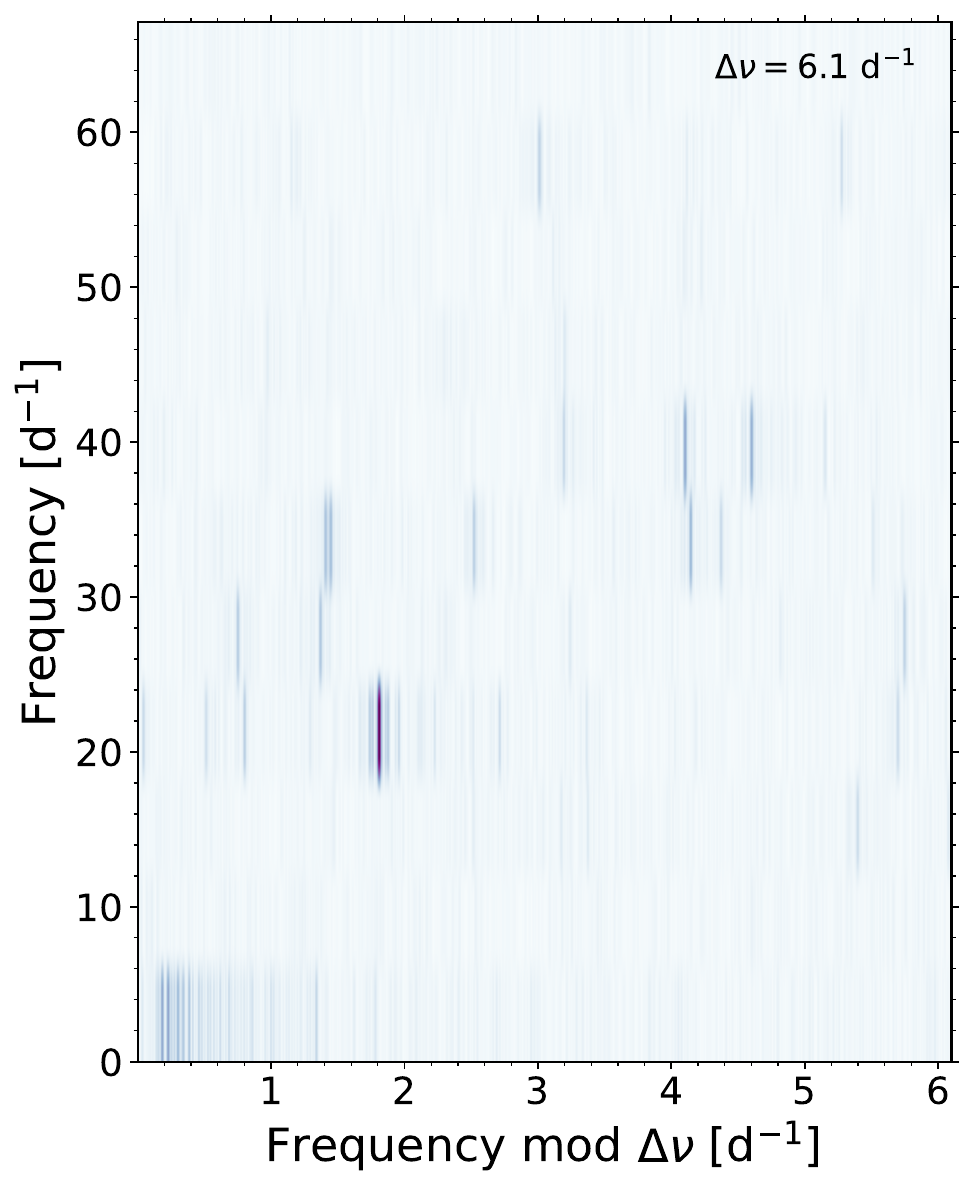}
        \caption{TIC~94724614}
        \label{echelle2}
    \end{subfigure}

    \begin{subfigure}{0.48\textwidth}
        \centering
        \includegraphics[width=\linewidth]{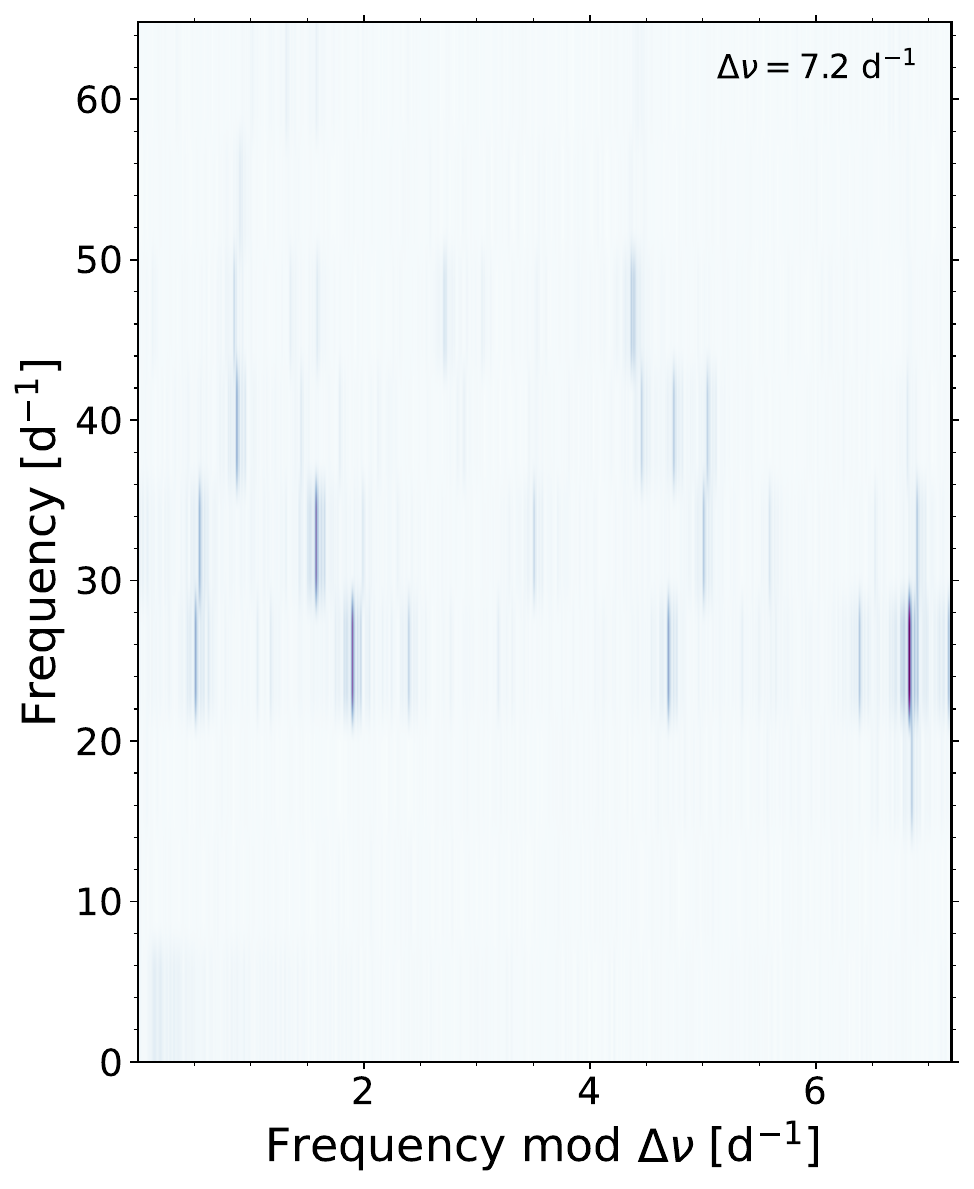}
        \caption{TIC~101168697}
        \label{echelle3}
    \end{subfigure}
    \hfill
    \begin{subfigure}{0.48\textwidth}
        \centering
        \includegraphics[width=\linewidth]{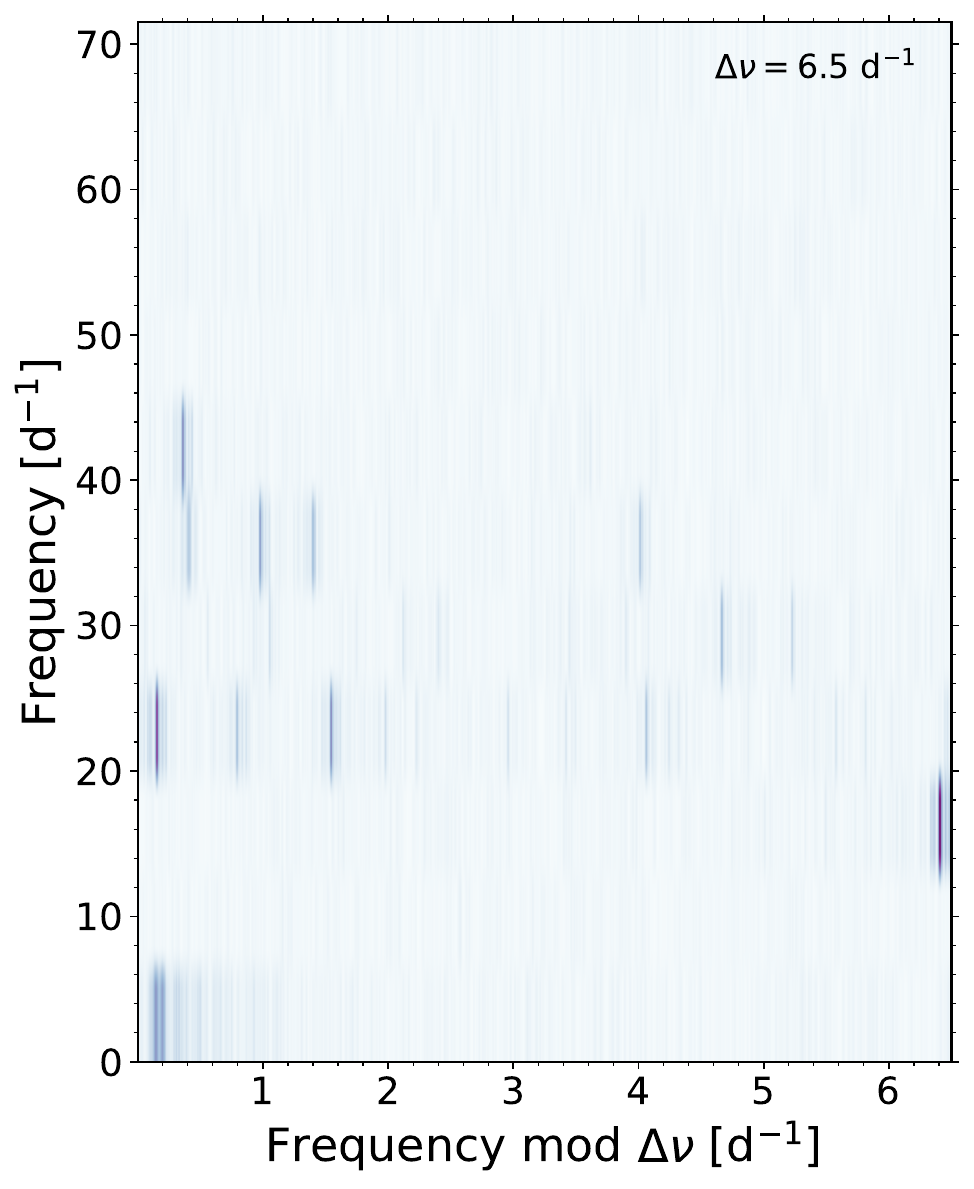}
        \caption{TIC~126098193}
        \label{echelle4}
    \end{subfigure}

    \caption{Échelle diagrams for the four $\delta$~Sct stars showing approximate regular frequency spacings. The adopted values of $\Delta\nu$ are indicated in each panel.}
    \label{echelle}
\end{figure*}

\begin{figure*}
   \centering
   \includegraphics[width=\columnwidth]{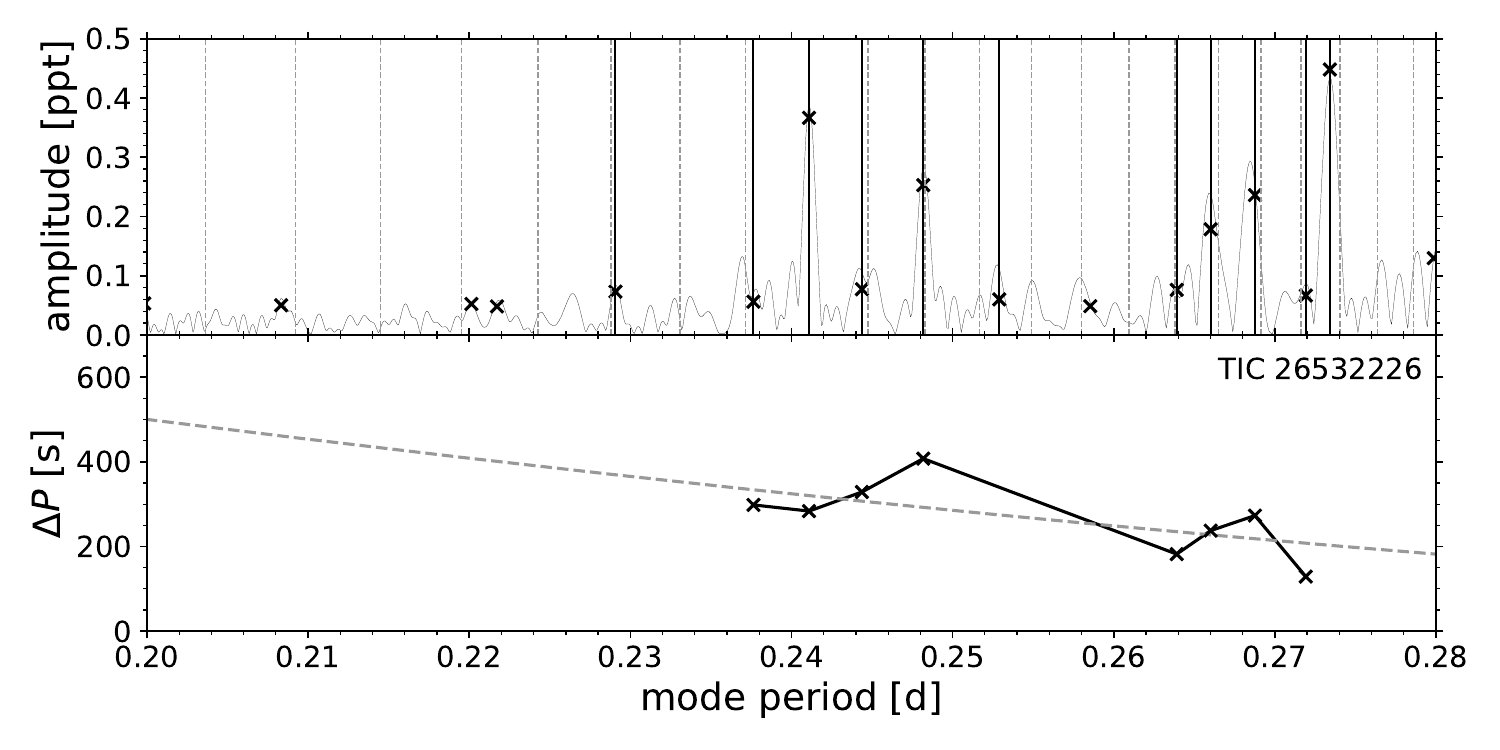}
    \includegraphics[width=\columnwidth]{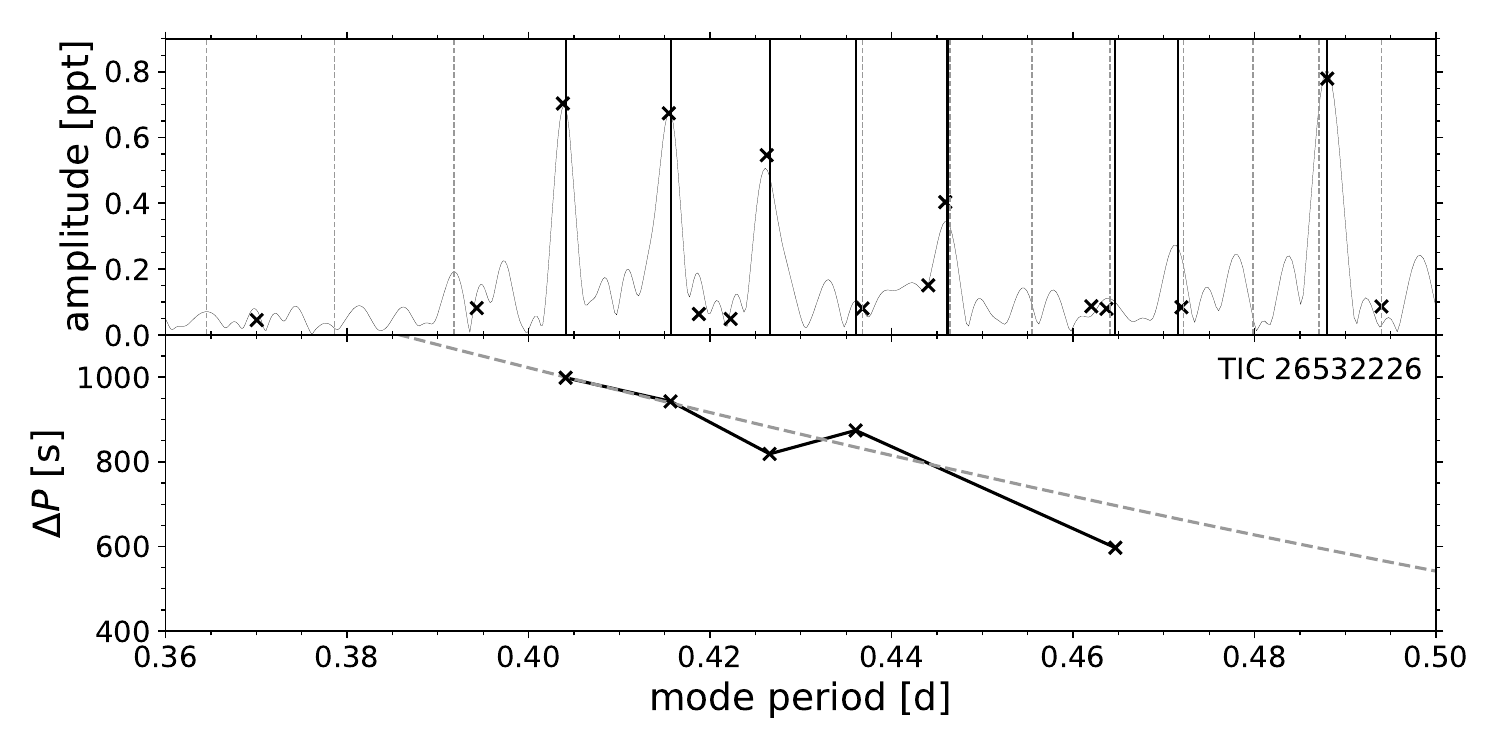}
    \caption{Independent fits to the prograde quadrupole (left) and dipole (right) period-spacing patterns of HD~281229. Plotting conventions follow Fig.~\ref{sim_psp}.}
\end{figure*}

\begin{table*}
    \centering
    \caption{Manually masked time intervals in the TESS light curves. All times are given in Barycentric TESS Julian Date (BTJD).}
    \label{masked_intervals}
    \begin{tabular}{cc}
        \hline \hline
        Sector & Masked interval \\
        \hline
        42 & 2457--2462 \\
        43 & 2484--2489 \\
        44 & 2510--2515 \\
        58 & 2894--2899 \\
        70 & 3219--3223 \\
        71 & 3246--3249 \\
        \hline
    \end{tabular}
\end{table*}

\begin{table*}
\centering
\caption{Rotational variables identified in the sample.}
\label{rotators}
\begin{tabular}{ccccc}
\hline
\hline
TIC ID & \emph{Gaia} DR3 ID & \emph{Gaia} G magnitude & $G_{\mathrm{BP}}-G_{\mathrm{RP}}$ & Rotation rate \\
&&[mag]&[mag]& [d$^{-1}$] \\
\hline
26653246  & 218804811091303936 & 5.727925  & 0.000590 & 1.047 \\
467643954 & 170139052056184320 & 8.554622  & 0.210091 & 1.465 \\
467651777 & 169262363331779328 & 8.595701  & 0.304214 & 1.292 \\
94473447  & 168851180342236544 & 8.603040  & 0.340007 & 2.487 \\
138115468 & 123999730221048320 & 9.110842  & 0.553643 & 3.612 \\
94335799  & 219798494725179008 & 9.411253  & 0.192955 & 1.467 \\
94279142  & 218807078833972224 & 9.551595  & 0.192775 & 1.060 \\
26649326  & 168239783157120768 & 9.577903  & 0.608263 & 0.615 \\
385555521 & 72049489436410752  & 10.001793 & 0.345196 & 1.510 \\
26392157  & 168137253698063488 & 10.248975 & 0.859394 & 0.986 \\
126098356 & 216708248573262208 & 10.661647 & 1.688691 & 0.164 \\
467651557 & 170024977725225600 & 10.720671 & 0.712235 & 0.288 \\
101166990 & 216531983114472960 & 10.866867 & 1.200633 & 0.229 \\
346677767 & 217358644060557696 & 11.141840 & 1.199107 & 3.932 \\
26355640  & 216715498478030720 & 11.439731 & 0.926010 & 0.560 \\
101255990 & 216681207459195392 & 11.626451 & 1.121111 & 0.270 \\
467651347 & 170063976028209536 & 11.763729 & 1.020580 & 0.230 \\
101220387 & 217444440327486848 & 13.440225 & 1.995058 & 0.258 \\
408195683 & 217475123573808000 & 13.567371 & 2.139091 & 1.181 \\
417832771 & 120934532320922496 & 13.817296 & 2.453363 & 0.703 \\
101166999 & 216531914394996864 & 13.971604 & 2.328115 & 0.966 \\
101255738 & 216721545790356864 & 11.384773 & 1.314075 & 3.980 \\
94371830  & 218917442313787776 & 14.564531 & 2.197319 & 0.420 \\
26393425  & 216634753093004416 & 14.838723 & 2.563672 & 0.598 \\
641805418 & 218946609436835712 & 10.754076 & 0.877868 & 0.632 \\
26499238  & 216427598231002752 & 15.382523 & 2.488544 & 0.226 \\
94279293  & 218791376433618304 & 14.321853 & 1.904540 & 0.674 \\
126098176 & 216629874010075776 & 14.588853 & 2.342612 & 0.210 \\
94476762  & 220114088922109824 & 15.561749 & 2.374375 & 1.090 \\
94545660  & 170394207473116160 & 14.296881 & 2.279813 & 0.371 \\
641805419 & 218946613727945344 & 9.865675  & 0.243207 & 0.123 \\
101255965 & 216681581120776576 & 14.425531 & 2.258114 & 0.494 \\
385555553 & 72051104344106240  & 15.236897 & 2.567381 & 0.849 \\
417832644 & 121009058593358848 & 14.020135 & 2.416115 & 0.428 \\
\hline
\end{tabular}
\end{table*}

\begin{sidewaystable*}
\caption{$\nu_{\rm max}$ estimates for $\delta$~Sct stars obtained using different methods.}
\label{tab:numax_methods}
\centering
\small
\begin{tabularx}{0.95\textwidth}{l *{8}{>{\centering\arraybackslash}X}}
\hline\hline
TIC ID & \emph{Gaia} DR3 ID & $T_\mathrm{eff}$ & \emph{Gaia} G magnitude & Amplitude Moment & Power Moment & Amplitude Smoothing & Power Smoothing & Mean $\nu_{\rm max}$ \\
       & &[K]&[mag]&[d$^{-1}$] & [d$^{-1}$]   & [d$^{-1}$] & [d$^{-1}$]  & [d$^{-1}$] \\
\hline
94724614  &168522907401977472
&8081&10.03
&30.63&28.49&29.06&29.41&29.39 \\
26391048  &216648703146774016
&9616&9.60
&49.04&49.10&48.57&49.12&48.96 \\
94724004  &170106101067203968
&7255&11.09
&20.85&20.14&22.85&21.37&21.30 \\
101091831 &121243014052056320
&7442&9.31
&43.09&45.57&42.78&46.00&44.36 \\
101129521 &217313117407234944
&--&13.71
&55.83&56.22&55.15&53.16&55.09 \\
101168697 &216565827456753152
&7480&10.20
&32.95&30.25&32.46&30.23&31.47 \\
126098193 &216630144591496832
&7323& 12.66
&27.04&25.79&29.22&27.37&27.36 \\
101166716 &217344041171754368&7752&11.94
&26.51&25.90&25.79&27.77&26.49 \\
\hline
\end{tabularx}
\end{sidewaystable*}

\end{document}